\documentclass[12pt]{amsart}
\usepackage{amsmath}
\usepackage{amssymb}
\usepackage{bbm,bm,dsfont}
\usepackage{inputenc}
\usepackage[T1]{fontenc}
\newcommand{\ip}[2]{\left\langle\,#1\,|\,#2\,\right\rangle}
\newcommand{\mbs}{\mathbf S}
\newcommand{\mbl}{\mathbf L}
\newcommand{\mbo}{\mathbf O}
\newcommand{\wmbo}{\widetilde{\mbo}}
\newcommand{\mbe}{\mathbf E}
\newcommand{\C}{\mathbb C}
\newcommand{\Q}{\mathbb Q}
\newcommand{\R}{\mathbb R}
\newcommand{\quat}{\mathbb H}
\newcommand{\br}{\mathcal B(\mathbb R)}
\newcommand{\hA}{\mathcal{A}}
\newcommand{\hB}{\mathcal{B}}
\newcommand{\hO}{\mathcal{O}}
\newcommand{\tr}[1]{\mathrm{tr}\bigl[#1\bigr]}
\newcommand{\sfe}{\mathsf{E}}
\newcommand{\sff}{\mathsf{F}}
\newcommand{\sfm}{\mathsf{M}}
\newcommand{\hi}{{\mathcal H}}
\newtheorem{axiom}{Axiom}

\newcommand{\mbf}{{\mathbf F}}
\newcommand{\pmbe}{{\rm ex}(\mbe)}
\newcommand{\ps}{\mathbf P}
\newcommand{\atl}{{\rm At}(\mbl)}
\newcommand{\ov}{\overline{v}}
\newcommand{\q}{\mathbf q}
\newcommand{\p}{\mathbf p}
\newcommand{\bv}{\mathbf v}

\newcommand{\sat}{\mathbf S_{\rm at}}
\newcommand{\esex}{(\sfe,X)}

\newtheorem{lemma}{Lemma}
\newtheorem{remark}{Remark}
\newtheorem{theorem}{Theorem}
\newtheorem{proposition}{Proposition}
\newtheorem{definition}{Definition}
\newtheorem{corollary}{Corollary}

\newtheorem{example}{Example}

\usepackage{color}

\begin{document}

\title{An Axiomatic Basis for Quantum Mechanics}

\author{Gianni Cassinelli}
\address{Department of Physics, University of Genova, and INFN Sezione di Genova, Genoa, Italy}
\email{cassinelli@ge.infn.it} 

\author{Pekka Lahti}
\address{Turku Centre for Quantum Physics, Department of Physics and Astronomy, University of Turku, Turku, Finland}
\email{pekka.lahti@utu.fi}

\begin{abstract}
In this paper we use  the framework of generalized probabilistic theories  to present two sets of basic assumptions, called axioms, for which we show that they lead to the Hilbert space formulation of quantum mechanics. The key results in this derivation are the co-ordinatization of generalized geometries and a theorem of Sol\' er which characterizes Hilbert spaces among the  orthomodular spaces. 
A generalized Wigner theorem is applied to reduce some of the assumptions of Sol\' er's theorem to the theory of symmetry in quantum mechanics. Since this reduction is only partial we also point out the remaining open questions.

\end{abstract}

\maketitle

\section{Introduction}

This paper  aims to give an overview of the axiomatic basis of quantum mechanics.
We  show that quantum mechanics on Hilbert space can 
to a large extent  
be derived from physically motivated assumptions  using either the 
quantum logic approach  or the  convexity approach,
both being examples of general probabilistic theories. The key results in these derivations are the coordinatization of generalized geometries, 
a theorem of Sol\' er, 
and a generalized Wigner theorem.
We also point out a mathematical assumption, which seems unavoidable but still lacks an operational justification.

The historic paper \cite{BvN1936} of Birkhoff and von Neumann, entitled {\em  The logic of quantum mechanics},
marks the beginning of the investigations on the mathematical and conceptual foundations of quantum mechanics which go under the title quantum logic. The literature of the field is very rich. In addition to the influential lecture notes of Mackey \cite{Mackey1963} we mention here only some  representative monographs to indicate the diversity of
the field \cite{EnricoGianni,Gudder1979,Jauch1968,Mittelstaedt1978,Piron1976,PavelSylvia,Redei1998,Varadarajan1985}.

The papers of Ludwig \cite{Ludwig1967}, Mielnik \cite{Mielnik1969,Mielnik1974}, Davies and Lewis \cite{DaviesLewis1970}, and Edwards \cite{Edwards1970,Edwards1971,Edwards1972}
have strongly influenced the development of the convexity or operational approach to quantum mechanics.
In addition to the mentioned original papers, the monumental work of Ludwig \cite{Ludwig1983} as well as  the monographs \cite{HartkamperNeumann1974,Davies1976}
are valuable sources for the physical and mathematical ideas behind the convextity approaches. 
A closely related approach is the empirical logic framework developed by Foulis and Randall
to study the manuals of physical operations; see, for instance, their papers \cite{FoulisRandall1973,FoulisRandall1974}.

In recent years quantum information theory has renewed interest in the foundations of quantum mechanics and some
new ideas have been proposed  for an axiomatic foundations of quantum mechanics,
see, e.g.,  \cite{Chiribellaetal2010,Chiribellaetal2011,Clifton_etal2003} and the many references given therein.  
Though interesting, these investigations deal only with the so-called finite level systems. 
Finite level systems are only a  part of quantum mechanics, leaving open 
the more general question on the derivability of quantum mechanics in infinite dimensional Hilbert space, which is needed, for instance, if one assumes that physical systems exist in four dimensional spacetime $\R^4$.
Our aim is to investigate this problem.

The structure of the paper is the following. The first part of the paper, Sections~\ref{sec:statistical duality} - \ref{sect:filters}, discusses a general probabilistic framework for an axiomatic basis of quantum mechanics. 
Section~\ref{sec:statistical duality} reviews the  idea of statistical duality, the concepts of states, experimental functions, operations, and effects, and  defines the basic structures.  We summarize them in two pairs of axioms, the weaker ones \ref{set_of_states} and \ref{affine}, and  the stronger ones \ref{statespace} and   \ref{effects}, concerning the sets of states $\mathbf S$ and experimental functions $\mathbf E$. 
In Section~\ref{sec:quantum logic} quantum logic  is defined as a pair  $(\mathbf S,\mbl)$,  $\mbl\subset\mathbf E$, arising from the structure $(\mathbf S,\mathbf E)$ of Axioms  \ref{set_of_states} and \ref{affine}
as specified further through the orthogonality postulate (Axioms~\ref{axiom1})  and an axiom which stipulates the
existence of a sufficiently rich family $\ps$ of pure states, extremal elements  in $\mathbf S$, expressed as probability measures on $\mbl$ 
(Axiom~\ref{axiom2}).
After excluding the classical case in Section~\ref{sect:classicalexcluded}, Corollary~\ref{AClattice1} concludes that the
fundamental structure attached to a proper quantum system consists of
$\mbl$ being an irreducible complete orthomodular AC lattice where the atoms of $\mbl$
are in a bijective correspondence with the pure states in $\mbs$.
As an alternative approach, Section~\ref{sect:filters} starts with the stronger pair of axioms, \ref{statespace} and \ref{effects},
defining a subset of operations, called filters, together with a subset of experimental functions, called propositions, and poses the projection postulate, Axiom~\ref{PP}, to build a one-to-one onto connections between the sets of filters and propositions. This leads, once again, to the structure  of Corollary~\ref{AClattice1}.

The two final Sections~\ref{HScoordinatization} - \ref{sec:symmetries} constitute the second part of the paper. 
There we study the problem of realizing the abstract structure $(\mathbf S,\mathbf E)$, with the substructure $(\ps,\mbl)$ of
Corollary~\ref{AClattice1}, as the one given by the Hilbert space formulation of quantum mechanics. In  this realization,
$\mathbf E$ is identified with the set of effect operators (positive unit bounded operators),
$\mbl$ as the extremal elements of $\mathbf E$, that is, the projection operators,  $\mathbf S$, via Gleason's theorem, as the density operators  (positive trace one operators), and $\ps$ as the extremal elements of $\mathbf S$, the one-dimensional projections, with $\tr{\rho E}$  giving the probability for an effect $E$ in state $\rho$.

To obtain the Hilbert space realization of $(\mathbf S,\mathbf E)$ we follow the somewhat indirect, and presumably not  optimal, method that starts by identifying $\mbl$ with the lattice of the closed subspaces of an orthomodular space  over a division ring, Sect.~\ref{ssect:reptheorem}. 
Exhibiting an example of a finite level non-Hilbertian model with the structure of Corollary~\ref{AClattice1},
we proceed to apply a theorem of Sol\'er (Sect.~\ref{ssect:soler}) to fix the orthomodular space to be an infinite dimensional  (real, complex, or quaternionic) Hilbert space.
 The final Section~\ref{sec:symmetries}
reduces the assumptions required by Sol\'er's theorem to the idea of symmetry, indicating, at the same time an unavoidable assumption whose physical meaning remains yet to be clarified. The choice between real, complex and quaternionic cases is briefly discussed at the end of the paper.

\section{Statistical duality and its representation}\label{sec:statistical duality}
\subsection{Statistical duality}

A general probabilistic formulation of a physical theory builds on the concepts of states  and observables and on the idea of statistical causality merging the two entities into a probability measure. {\em States} are understood as equivalence classes of preparations of a physical system,
{\em observables} as equivalence classes of measurements on it, and the {\em statistical causality} claims that any state $\alpha$ and any observable $\sfe$ determine a probability measure  $p(\alpha,\sfe,\cdot)$ defined on a $\sigma$-algebra $\hA$ of subsets of a (nonempty) set $\Omega$,
with $p(\alpha,\sfe,X)$ giving the probability that a result is registered in the set $X$ when a measurement of $\sfe$ is performed on the system prepared in state $\alpha$. 
An observable $\sfe$ thus goes with a value space $\Omega$ together with the test sets $X\in\hA$ within which the results are counted. 
To emphasize this, we may also write $(\sfe,\Omega,\hA)$ for $\sfe$.
In most applications
$(\Omega,\hA)$ is just the real Borel space $(\R,\br)$, or a Cartesian product space $(\R^n,\mathcal{B}(\R^n))$, or a (Borel) subspace of such spaces.

Let $\mbs$ and $\mbo$ be the sets of all states and all observables of the system. We call the pair $(\mbs,\mbo)$ together with the probability function 
$p:(\alpha,\sfe)\mapsto p(\alpha,\sfe,\cdot)$
a {\em statistical duality}.
In an axiomatic approach one aims to  introduce physically motivated structures for the sets $\mbs$ and $\mbo$ so that the form of
the probability measures $p(\alpha,\sfe,\cdot)$, $\alpha\in\mbs, \sfe\in\mbo$, gets determined.

\subsection{State space}\label{sect:statespace}
\subsubsection{Convex structure}
The set $\mbs$ of states  can immediately be equipped with a convex structure reflecting the possibility of combining preparations (and thus states) into new preparations (and thus states) by statistically mixing them. 
Indeed, for any $\alpha,\beta\in\mbs$ and $0\leq \lambda\leq 1$ one may define a state $\langle\lambda,\alpha,\beta\rangle$ through the $\lambda$-convex
combination of the probability measures $p(\alpha,\sfe,\cdot)$ and $p(\beta,\sfe,\cdot)$, $\sfe\in\mbo$,
\begin{equation}\label{convexstructure}
p(\langle\lambda,\alpha,\beta\rangle,\sfe,\cdot)=\lambda p(\alpha,\sfe,\cdot)+(1-\lambda)p(\beta,\sfe,\cdot).
\end{equation}
The existence of such a state 
$\langle\lambda,\alpha,\beta\rangle$ is built in the assumption that preparations can be statistically mixed to produce new preparations and the uniqueness of  $\langle\lambda,\alpha,\beta\rangle$ follows from the statistical completeness built in the notions of states (as equivalence classes of preparations) and observables (as equivalence classes of measurements).

The existence of the function $[0,1]\times\mbs\times\mbs\to\mbs$ with the property (\ref{convexstructure}) defines what is known as a {\em convex structure}, and it is then 
a mathematical convenience to consider $\mbs$ as properly placed in a real vector space $U$ 
so that we may simply write $\langle\lambda,\alpha,\beta\rangle=\lambda\alpha+(1-\lambda)\beta$ \cite[Theorem 2]{Stone1949}, see also
\cite{Gudder1973, CornetteGudder1974}.

Using $\sigma$-convex combinations of probability measures one may also introduce $\sigma$-convex combinations of states $(\alpha_i)$ with
weights $(\lambda_i), \lambda_i\geq 0, \sum\lambda_i=1$, through $\sum\lambda_ip(\alpha_i,\sfe,\cdot), \sfe\in\mbo$, with the obvious requirement that the series 
$\sum\lambda_ip(\alpha_i,\sfe,X)$ is convergent for each $\sfe,X$.
If there is an $\alpha\in\mbs$
such that $p(\alpha,\sfe,\cdot)= \sum\lambda_ip(\alpha_i,\sfe,\cdot)$ for all  $\sfe\in\mbo$,  we say that $\alpha$ is a $\sigma$-convex
combination of the states $(\alpha_i)$ with weights $(\lambda_i)$ and we write $\alpha=\sum\lambda_i\alpha_i$. 
Again, if such a state exists it is unique. 
As seen from Theorem~\ref{EG1970} below it is a mathematically convenient idealization to assume that the set of states is also closed under $\sigma$-convex combinations.

The convex structure of $\mbs$ allows the distinction between the {\em pure states}, the extreme elements of $\mbs$, and the {\em mixed states}, the nonextreme elements of $\mbs$.
We let $\ps={\rm ex}(\mbs)$ denote the set of all pure states in $\mbs$. The existence of pure preparations and thus pure states is another natural assumption supported equally well by everyday
experience as well as by sophisticated quantum experiments.

We summarize the above discussion in  the first axiom.

\begin{axiom}\label{set_of_states}
The set of all states of a physical system 
described by the statistical duality  $(\mbs,\mbo,p)$ forms a convex subset of a real vector space. 
\end{axiom}

We specify later the assumptions concerning $\sigma$-convex combinations of states as well as the existence of a sufficiently large set of pure states.

\subsubsection{$\mbs$ as a base for a generating cone}
Let
$K=\{\lambda\alpha\,|\, \lambda\in\R^+, \alpha\in\mbs\}=\cup_{\lambda\geq 0}\lambda\mbs\subset U$ be the cone defined by $\mbs$.
We assume now that this cone is a proper cone, that is, $K\cap -K=\{0\}$, 
and that each $\gamma\in K$, $\gamma\ne 0$, has a unique representation as $\gamma=\lambda\alpha$ for some $\lambda>0$ and $\alpha\in\mbs$.
Let $V=K-K$ be the real vector space generated by $K$ (possibly a  subspace of $U$). Then
$K=V^+=\{v\in V\,|\, v\geq 0\}$ and $\mbs$ is a base of the cone $K$. 
The existence of a base for a generating cone of a real vector space $V$ is known to be equivalent with the existence of a strictly positive linear functional on $V$ \cite[Lemma 2]{Ellis1964}.
We
let  $e:V\to\R$ be the functional defined by $\mbs$ so that
\begin{equation}
\mbs =\{\alpha\in V^+\,|\, e(\alpha)=1\}
\end{equation}
and call it the  {\em intensity} functional.

The physical interpretation attached to $\mbs$ can be extended to the positive cone $V^+$: any $\lambda\alpha$, $\lambda\in\R^+,\alpha\in\mbs$, represents a new state of the system obtained from $\alpha$ by changing its intensity. The elements of $\mbs$ will be distinguished as normalized states, whereas the term `state' will be extended to refer to all elements of $V^+$, including the empty state, the null element of $V$. The linear operations $(\alpha,\beta)\mapsto\alpha+\beta$ and $(\lambda,\alpha)\mapsto\lambda\alpha$, $\alpha,\beta\in V^+,\lambda\in\R^+$, preserve their original interpretation as mixing and intensity changing, respectively. In particular, the term
{\em pure state} can thus refer to an element of ${\rm ex}(\mbs)$ or of ${\rm ed}(V^+)=\{\lambda\alpha\,|\, \lambda\in\R^+,\alpha\in{\rm ex}(\mbs)\}$, where ${\rm ed}$ stands for `edge'.
This extension will be accepted only as a mathematically convenient way of speaking and it has no physical implications.

Let ${\rm conv}(\mbs\cup-\mbs)$ denote the convex hull of the set $\mbs\cup-\mbs$. 
This set is 
 convex,  absorbing (for any $v\in V$ there is a $\lambda>0$ such that $v\in\lambda{\rm conv}(\mbs\cup-\mbs)$), and balanced ($\lambda v\in{\rm conv}(\mbs\cup-\mbs)$ for all $v\in{\rm conv}(\mbs\cup-\mbs)$ and $-1\leq\lambda\leq 1$).
Therefore its Minkowski functional (or gauge)
$p_\mbs:V\to\R$, defined as
$$
p_\mbs(v)=\inf\{\lambda>0\,|\, v\in\lambda {\rm conv}(\mbs\cup-\mbs)\},
$$
is a seminorm \cite[Theorem II.1.4]{Schaefer1971}. 
Clearly, $p_\mbs(\alpha)=e(\alpha)$ for all $\alpha\in V^+$ and  we note that for all $v\in V$
$$
p_\mbs(v)=\inf\{e(\alpha)+e(\beta)\,|\, \alpha,\beta\in V^+, v=\alpha-\beta\}.
$$
If this seminorm is a norm, then $(V,V^+,\mbs)$ or just $(V,\mbs)$  is a {\em base norm space}. The following result, due to Edwards and Gerzon \cite{EdwardsGerzon1970} is now crucial:

\begin{theorem}\label{EG1970} If the set $\mbs$ of all states of the system forms a base for a generating cone of a vector space $V$ and
is $\sigma$-convex, then its Minkowski functional $p_\mbs:V\to\R$ is a norm 
with respect to which $V$ is a Banach space.
\end{theorem}

There is another technical point to be noted. According to a proposition of  Ellis \cite{Ellis1966},
if $(V,V^+,\mbs)$ is a base norm space, then also $(V,\overline{V^+},\overline\mbs)$ 
(where $\overline{A}$ denotes the norm closure of
a subset $A$ of $V$)
is a base norm space and its norm coincides with the norm of $(V,\mbs)$. We note, in addition, that $\mbs$ is closed if and only if $V^+$ is closed. 
With the risk of adding some nonphysical elements in the set $\mbs$ we now 
formulate an alternative  stronger assumption concerning the set of states of a statistical duality,  as first formulated in \cite{DaviesLewis1970}.

\begin{axiom}\label{statespace}
The set of all states of a physical system 
described by the statistical duality $(\mbs,\mbo,p)$
is represented by a norm closed generating cone $V^+$ of a base norm Banach space $(V,\mbs)$.
\end{axiom}

We have formulated two different axioms concerning the basic assumptions on the set of states,
the weaker axiom serving as a starting point for the quantum logic approach, Section \ref{sec:quantum logic}, 
the stronger axiom defining the beginning of the convexity or state space approach, Section \ref{sect:filters}.

\subsection{Experimental functions}
\subsubsection{Affine maps $\mbs\to[0,1]$}

With every observable $(\sfe,\Omega,\hA)\in\mbo$ one may consider  the family of all ordered pairs $\langle\sfe,X\rangle, X\in\hA$,
called experimental pairs. With each such pair we can associate the statement
`a measurement of $\sfe$ yields a result in the set $X$', denoted by $\esex$ and called an {\em experimental statement}. 
Then the number $p(\alpha,\sfe,X)$, $\alpha\in\mbs$ gives the probability  for the statement $\esex$ to be true in the state $\alpha$.
Experimental statements $\esex$ and $(\sff,Y)$
are said to be {\em equivalent} if for all $\alpha\in\mbs$,
\begin{equation}\label{M3}
p(\alpha,\sfe,X)=p(\alpha,\sff,Y).
\end{equation}
This defines an equivalence relation 
in the set of all experimental statements $\{\esex\,|\,\sfe=(\sfe,\Omega,\hA)\in\mbo, X\in\hA\}$.
Let $\mbe_0$
denote the set of all equivalence classes $|\esex|$ of the statements $\esex$.
A given  element of $\mbe_0$ is denoted by a letter $a$ and is called an {\em (experimental) proposition}. 
The experimental proposition $a$ is a set of experimental statements equivalent among themselves with respect to $p$.

There is a fundamental difference between $\esex$ and $|\esex|$. Namely, $\esex$ is nothing more than a statement saying that a measurement of $\sfe$ yields a result in $X$, and it does not depend on $p$; in contrast, $|\esex|$  represents the proposition  that every two statements $(\sfe_1,X_1)$ and $(\sfe_2,X_2)$ from $|\esex|$ are equivalent to $\esex$, that is, for every $\alpha\in\mbs$,  $p(\alpha,\sfe_1,X_1)=p(\alpha,\sfe_2,X_2)=p(\alpha,\sfe,X)$. The experimental proposition $|\esex|$ clearly depends on $p$; 
it should be written more exactly $|\esex|_p$. Hence to specify the proposition $|\esex|_p$ one has to measure of all the experimental statements 
$(\sff,Y)$ and find  those   equivalent  to $\esex$. Hence the experimental proposition $a=|\esex|_p$ is a subset of all experimental statements depending on $p$, typically much bigger than the one-element set $\{\esex\}$.

Each experimental proposition $a\in\mbe_0$ defines a real 
valued function, {\em experimental function}
\begin{equation}\label{M4}
f_a:\mbs\to[0,1],\  f_a(\alpha)=p(\alpha,\sfe,X),\quad (\sfe,X)\in a,
\end{equation}
and we let $\mbe\subset [0,1]^{\mbs}$ denote the set of all such functions. 
 If $f\in\mbe$, then $f(\alpha)$, $\alpha\in\mbs$,
is the probability that the proposition $a=|(\sfe,X)|$, with $f=f_a$,  is true in state $\alpha$, that is, an $\sfe$-measurement in state 
$\alpha$ yields a result in $X$. 

Consistently with the very definition \eqref{convexstructure} of mixed states it is natural to assume that the experimental functions are affine, that is, we have the following axiom.

\begin{axiom}\label{affine}
The set $\mbe$ of experimental functions of a statistical duality $(\mbs,\mbo,p)$ is a subset of the set of affine functions $\mbs\to[0,1]$.
\end{axiom}

We let $0$ and $e$ denote the constant zero and one functions $\mbs\to[0,1]$, respectively.
Clearly, $0,e\in\mbe$, and if $f\in\mbe$ then also $f^\perp=e-f\in\mbe$ so that
for any $\alpha\in\mbs$, $f(\alpha)+f^\perp(\alpha)=1$. 
Moreover, as real valued functions,
the set $\mbe$ is partially ordered in a natural way, 
that is, for any $f,g\in\mbe$, $f\leq g$ if and only if $f(\alpha)\leq g(\alpha)$ for all $\alpha\in\mbs$, the operational content being given by \eqref{M4}.

The set $\mbe$ has the order bounds 0 and $e$  
and the mapping $\mbe\ni f\mapsto f^\perp\in\mbe$ is an order reversing involution. However, it is not an orthocomplementation, that is,
the greatest lower bound of  a pair $(f,f^\perp)$ need not be 0.

Any observable $\sfe$ can now be represented as an $\mbe$-valued set function $\sfe:X\mapsto f_{|(\sfe,X)|}$ 
such that for each $\alpha\in\mbs$, the set function $X\mapsto  \alpha(f_{|(\sfe,X)|})=f_{|(\sfe,X)|}(\alpha)=p(\alpha,\sfe,X)$
is a probability measure. Moreover, any $f\in\mbe$ is in the range of some observable $\sfe$.
In this sense the set $\mbo$ of all observables is a {\em surjective} set of $\mbe$-valued set functions.

We shall follow two distinct approaches to specify further the structure on $\mbe$.
For that end, we already pose the following two definitions:

\begin{definition}\label{Maciej1}
A sequence (finite or countably infinite) of experimental functions $f_1,f_2,\ldots$  is  {\em orthogonal}
if there is an experimental function $g$ such that $g+f_1+f_2+\ldots=e$.
\end{definition}

\begin{definition}\label{Maciej2}
A sequence (finite or countably infinite) of experimental functions $f_1,f_2,\ldots$ is  {\em pairwise orthogonal}
if $f_i+f_j\leq e$ for all $i\ne j$, $i,j=1,2,\ldots$
\end{definition}

An orthogonal sequence is pairwise orthogonal but in general not conversely. 
Still  the concepts of orthogonality and pairwise orthogonality
 have the following common property: if for some state $\alpha$ one of the experimental statements $f_i$ is true, that is, $f_i(\alpha)=1$, then all the other experimental statements are false in that state, that is, $f_j(\alpha)=0$ for all $j\ne i$. This supports the hypothesis that the statistical duality $(\mbs,\mbo,p)$ might have a substructure where the two notions coincide; we return to that in Sect.~\ref{sec:quantum logic}.

\subsubsection{Positive unit bounded functionals on $V$}
Further properties of experimental functions can be obtained under the assumption of
Axiom \ref{statespace}. Indeed, in this case
any $f\in\mbe$ has a unique extension to a positive continuous linear functional on $V$ bounded by $e$. We denote this extension with the same symbol $f$. 
In this case, the set $\mbe$ of experimental functions is thus a subset of the order interval $[0,e]$ of the order unit Banach space $(V^*,[0,e])$. 
With the risk of adding some new  elements in the set $\mbo$ of all observables one could assume that actually $\mbe=[0,e]$.\footnote{Clearly, this assumption could also be posed under Axiom \ref{set_of_states} but we refrain of doing it.}
We express also this stronger assumption concerning experimental functions as a further possible axiom.
\begin{axiom}\label{effects}
The set $\mbe$ of all experimental functions coincides with the order interval $[0,e]$ of the dual Banach space of $(V,\mbs)$.
\end{axiom}

This axiom has a simple but important consequence: for any two $f,g\in\mbe$, if $f\leq g^\perp$, then also $f+g\in\mbe$. 
Clearly, then $f,g\leq f+g$, but this does not mean that their smallest upper bound $f\lor g$ would exist in $\mbe$, and even if it would exist, it need not equal to $f+g$.

Another important structure of the set $\mbe$ arising from Axiom~\ref{effects} is its convexity; for any $f,g\in\mbe$ and $0\leq\lambda\leq 1$, $\lambda f+(1-\lambda)g\in\mbe$.
Since the order interval $\mbe\subset V^*$ is also compact (by the Banach-Alaoglu theorem),
the Krein-Milman theorem says that the closure of the convex hull of the extremal elements of $\mbe$ is the whole set of experimental functions, that is, $\overline{{\rm conv}}({\rm ex}(\mbe)) = \mbe$. 

\subsection{Operations}\label{sect:operationseffects}
The number $p(\alpha,\sfe,X)=f_{|(\sfe,X)|}(\alpha)$ is the probability that a measurement of $\sfe\in\mbo$ in the state $\alpha\in\mbs$ leads to a result in the set $X\in\hA$.
Such a  measurement may destroy the system or, in any case, 
cause a change in its state. In addition to such a forced change the system may also experience a spontaneous change, for instance, in 
the course of its time evolution.

To describe such state changes we now build on axioms \ref{statespace} and \ref{effects}.
It is also convenient to allow the possibility that the intensity of a state may change in the process:
 $V^+\ni\alpha\mapsto\alpha'\in V^+$, including the possibility
that $\alpha'=0$, that is, the system gets destroyed in the intervention. We consider only such changes on the system which can be described by functions $V^+\ni\alpha\mapsto\phi(\alpha)\in V^+$, with the 
obvious interpretation that $\alpha$ is the state of the system before the change and
$\phi(\alpha)$ its state after the change. Various types of state changes $\alpha\mapsto\phi(\alpha)$ may occur in realistic physical
situations.

We restrict our consideration only to such state changes where the intensity of the state is not increasing, that is, we assume that for each state $\alpha\in V^+$,
\begin{equation}\label{op1}
e(\phi(\alpha))\leq e(\alpha).
\end{equation}
Consider then a mixed state $\beta=\lambda_1\alpha_1+\lambda_2\alpha_2$, with $\lambda_1,\lambda_2\in\R^+$ and $\alpha_1,\alpha_2\in V^+$. In a change $\phi$
the state $\beta$ transforms to $\phi(\beta)$ whereas the states $\alpha_1,\alpha_2$ transform to $\phi(\alpha_1)$ and $\phi(\alpha_2)$ of which one may form the
mixture $\lambda_1\phi(\alpha_1)+\lambda_2\phi(\alpha_2)$. There are physical situations where the state $\phi(\lambda_1\alpha_1+\lambda_2\alpha_2)$ may differ from the state
$\lambda_1\phi(\alpha_1)+\lambda_2\phi(\alpha_2)$.\footnote{The paper of Mielnik \cite{Mielnik1969} contains  an extensive analysis of possible state changes, including some nonlinear processes.} Again, we restrict our attention only to such changes $\phi$ for which these states are always the same, that is, for any
$\lambda_1,\lambda_2\in\R^+,\alpha_1,\alpha_2\in V^+$,
\begin{equation}\label{op2}
\phi(\lambda_1\alpha_1+\lambda_2\alpha_2)= \lambda_1\phi(\alpha_1)+\lambda_2\phi(\alpha_2).
\end{equation}
It is again a simple exercise to check that any map $\phi:V^+\to V^+$ with the properties (\ref{op1}) and (\ref{op2}) has a unique extension to a positive linear contracting mapping of
$V$ into $V$. We denote this extension by the same letter $\phi$ and we call such mappings {\em operations}.

Let $\bm{\hO}$ be the set of all operations. 
The sequential application of any two operations 
defines a new operation 
giving  $\bm{\hO}$ the structure of a noncommutative semigroup. 
Another physically relevant structure of $\bm{\hO}$ is that of  convexity, for any two $\phi_1,\phi_2\in\bm{\hO}$ and for any $0\leq \lambda\leq 1$, $\lambda\phi_1+(1-\lambda)\phi_2\in\bm{\hO}$,
which allows one to single out the extremal operations, that is, operations that cannot be obtained as  nontrivial mixings of any other operations. 

An operation $\phi\in\bm{\hO}$, when combined with the intensity functional $e$, defines an experimental function
$e\circ\phi\in\mbe$.
On the other hand, if $f\in\mbe$, 
then
fixing a $\beta\in\mbs$ and defining $\phi(\alpha)=f(\alpha)\beta$, $\alpha\in V$, one observes that $\phi\in\bm{\hO}$ and $e\circ\phi=f$.
The set of functionals $e\circ\phi,\phi\in\bm{\hO}$, 
thus coincides with the set of experimental functions. 
Due to this coincidence, the experimental functions are also called {\em effects}: $f\in\mbe$ is the effect of any operation $\phi\in\bm{\hO}$ such that $f=e\circ\phi$.
Extremal elements of $\mbe$ are called {\em decision effects} or {\em sharp} effects; hence, an effect is either sharp (extremal)  or unsharp (non-extremal).

Let $f$ be a decision effect. If $f=e\circ(\lambda\phi_1+(1-\lambda)\phi_2)$, for some $\phi_1,\phi_2\in\bm{\hO}, \lambda\in[0,1]$, then $e\circ\phi_1=e\circ\phi_2$, that is, the operations $\phi_1$
and $\phi_2$ are {\em isotonic}. Actually, the relation $e\circ\phi=e\circ\psi$, $\phi,\psi\in\bm{\hO}$, defines an equivalence relation 
in $\bm{\hO}$ and one may immediately confirm that there is a one-to-one onto correspondence between the set of effects and the isotony classes of operations. 
We let $[\phi]_f$ denote the isotony class of operations $\phi\in\bm{\hO}$ associated with the effect $f\in\mbe$.

One may now define an {\em instrument} as an operation valued set function $\hA\ni X\mapsto \phi_X\in\bm{\hO}$ for which $X\mapsto e\circ\phi_X$ is an observable, that is,
$X\mapsto e(\phi_X(\alpha))$ is a probability measure for each $\alpha\in\mbs$. By definition, any instrument defines an observable, but 
the converse holds also:
any observable $\sfe$ arises from some instrument such that $\sfe(X) = e\circ\phi_X$.
Calling two instruments isotonic if they define the same observable one again has that the isotony classes of instruments are in one-to-one onto correspondence with the observables of the system.

Preparing the system in a state $\alpha\in\mbs$, acting on it by an operation $\phi\in\bm{\hO}$, and detecting the (probabilistic) effect $e\circ\phi\in\mbe$ comprises the main steps in
the operational approach built on the statistical duality of states and observables.
To specify further stuctures of the description $(\mbs,\bm{\hO},\mbe)$ 
one may proceed in many different ways by posing fadditional conditions on any of the sets $\mbs$, $\bm{\hO}$, or $\mbe$. 
The  remark below is an indication how to reach classical descriptions
out of this general probabilistic model. Our aim is to pose conditions which lead to quantum descriptions.

\begin{remark}\label{classicalcase}
The space $V$ is ordered by the cone $V^+$ defined by the base $\mbs$. If this order is a lattice order, then $\mbs$ is a (Choquet) simplex \cite{Alfsen1971}, 
a structure considered to be characteristic of classical descriptions. If $V=V^+-V^+$ is a vector lattice, then also its dual $V^*$ is a vector lattice. In this case also the order interval $\mbe$ is a lattice 
and the set ${\rm ex}(\mbe)$ of extremal effects is a Boolean lattice with $f\mapsto f^\perp=e-f$ as the orthocomplementation \cite{Schaefer1974} - another
characteristic of classical descriptions.
\end{remark}

In the next two chapters
we shall follow two different approaches to specify further
the statistical duality $(\mbs,\mbo,p)$.
We start with an approach based on axioms \ref{set_of_states} and \ref{affine}. In Section \ref{sect:filters} we build on the stronger axioms \ref{statespace}  and \ref{effects}.

\section{Quantum logic}\label{sec:quantum logic}
The Mackey approach to quantum logic can be viewed as a further specification of the structures arising from the statistical duality $(\mbs,\mbo,p)$
with assuming, in the first instance, the existence of a sufficiently large subset of observables $\wmbo\subseteq\mbo$ for which the order
structure of the resulting subset of experimental functions  gets sufficiently regular. In this section we discuss assumptions of this kind. We stress once more that  here we  assume only that the set of states is convex and we allow the possibility that
the set of experimental functions is  a subset of the set of affine functions $\mbs\to[0,1]$.

\subsection{Orthogonality postulate}\label{OMsection}

Any subset of the set  $\mbo$ of all observables defines the corresponding subsets of the sets $\mbe_0$ and  $\mbe$. 
The basic assumption of the Mackey approach to quantum logic (Mackey's \cite{Mackey1963} Axiom V) can now be restated as the requirement on the existence of
a  (nonempty) subset $\wmbo\subseteq\mbo$ 
such that in the resulting subset $\mbl\subseteq\mbe$ of the experimental functions
the two notions of Definitions~\ref{Maciej1} and \ref{Maciej2} coincide.
We call this assumption the {\em orthogonality postulate} and formulate it as a further  axiom.

\begin{axiom}\label{axiom1}\rm
The set $\mbo$ of observables 
of the statistical duality $(\mbs,\mbo,p)$
contains a (nonempty) subset $\wmbo$ such that in the resulting set $\mbl$ of
experimental functions
a sequence $f_1,f_2,\ldots\in\mbl$  is orthogonal (in $\mbl$) if and only if it is pairwise orthogonal (in $\mbl$).
\end{axiom}

This axiom has important implications in the order structure of the set $\mbl$. 
First of all, it guarantees that for any two mutually orthogonal
elements $f,g\in\mbl$ also $f+g\in\mbl$. 
Moreover, it implies that the map $f\mapsto f^\perp$ is an orthocomplementation and it turns
$(\mbl,\leq,{}^\perp)$ into an orthomodular $\sigma$-orthocomplemented partially ordered set, with $0,e\in\mbl$ as the order bounds.

Though obvious, we note that for any two $f,g\in\mbl$, the set of their lower (upper) bounds in $\mbl$ is smaller
than the corresponding set in $\mbe$. Therefore, $f\land g$ may exists in $\mbl$ without existing in $\mbe$.

\begin{theorem}\label{Mtheorem}(M\c aczynski,\cite{Maczynski1973})
Let $(\mbs,\mbo,p)$ be a statistical duality and let $\wmbo$ be a (nonempty) subset of $\mbo$ such that the associated set $\mbl$
of experimental functions satisfies  Axiom~\ref{axiom1}. 
The set $\mbl$ 
is  an orthocomplemented orthomodular $\sigma$-orthocomplete  partially ordered set 
with respect to the natural order of real functions 
and the complementation $f^\perp=e-f$.
\end{theorem}

\noindent
{\bf Proof.}
{ 
Clearly, $\mbl$ is partially ordered by $\leq$ and  $0,e\in\mbl$. With $f\in\mbl$ also $f^\perp=e-f\in\mbl$, and $\mbl\ni f\mapsto f^\perp\in\mbl$ is 
an order reversing involution. 

Let $f_1,f_2\in\mbl$ and assume that $f_1+f_2\leq e$. Then by Axiom~\ref{axiom1} $f=f_1+f_2\in\mbl$. To show that $f$ is the least upper bound of $f_1$ and $f_2$ in $\mbl$, assume that $g\in\mbl$ is such that $f_1\leq g$ and $f_2\leq g$.
Then also $f_1+g^\perp\leq e$ and $f_2+g^\perp\leq e$ and thus $f_1+f_2+g^\perp\in\mbl$ so that $f_1+f_2\leq g$, that is $f_1+f_2=f_1\lor_{\mbl}f_2$.
By induction one  shows that $f_1+\cdots+f_n=f_1\lor_{\mbl}\cdots\lor_{\mbl}f_n$ for any pairwise orthogonal set $\{f_1,\ldots,f_n\}\subset\mbl$.
Let $(f_i)$ be a sequence of mutually orthogonal elements in $\mbl$ so that by assumption $f=f_1+f_2+\ldots\in\mbl$. Clearly, $f_i\leq f$ for each $i$. Let $g\in\mbl$ be such that $f_i\leq g$ for all $i$. Since for any $n$, $f_1+\cdots+f_n=f_1\lor_{\mbl}\cdots\lor_{\mbl}f_n$ we thus have
$f_1+\cdots+f_n\leq g$ for any $n=1,2,\ldots$ and therefore $f=f_1+f_2+\ldots\leq g$. 

For any $f\in\mbl$, $f+f^\perp\leq e$ and thus $e=f+f^\perp=f\lor_{\mbl}f^\perp$. By de Morgan laws we also have  $f\land_\mbl f^\perp = 0$ for any
$f\in\mbl$. This concludes the proof that $\mbl$ is orthocomplemented and $\sigma$-orthocomplete. 

To show orthomodularity, we need to show that for $f\leq g$, $f,g\in\mbl$, one has $g=f\lor_{\mbl}(g\land_{\mbl}f^\perp)$. If $f\leq g$, then
$f\lor_{\mbl}g^\perp=f+g^\perp=f+(e-g)$ and $h=(f\lor_{\mbl}g^\perp)^\perp=g-f\in\mbl$. Hence $f+h=g\leq e$ and thus $f\lor_{\mbl}h=f+h$ so that
$f\lor_{\mbl}(g\land_{\mbl}f^\perp)= f\lor_{\mbl}(f\lor_{\mbl}g^\perp)^\perp=f+h=g$. \qed
}

\

We call $\mbl$ the logic of $p$.
Henceforth  we simply write $f\land g$ for $f,g\in\mbl$
instead of $f\land_{\mbl}g$, and similarly for $f\lor g$, whenever the meet (join) exists in $\mbl$.

\begin{remark}\rm 
Consider an $f\in\mbl$, $0\ne f\ne e$, and assume that $\lambda f\in\mbl$ for some $0<\lambda<1$. Since $\lambda f\leq f$, then
$\lambda f+(e-f)\in\mbl$, and hence $e-(\lambda f +(e-f))=(1-\lambda)f\in\mbl$. Since $\lambda f +(1-\lambda)f=f\leq e$, also $\lambda f$ and $(1-\lambda)f$ are pairwise orthogonal so that their sum should equal to their least upper bound in $\mbl$, which is a contradiction.
In particular, $\mbl$ is not convex. 
\end{remark}

Each observable $\sfe=(\sfe,\Omega,\hA)\in\wmbo$ determines a unique $\mbl$-valued measure $\sfm_\sfe:\hA\to\mbl$ defined by $\sfm_\sfe(X)=f_{|(\sfe,X)|}$. By  Theorem~\ref{Mtheorem}, $\sfm_\sfe$ is in fact  a $\sigma$-homomorphism implying, in particular, that $\sfm_\sfe(\hA)$ is a Boolean sub-$\sigma$-algebra of $\mbl$.
We identify $\sfm_\sfe$ with $\sfe$.

Each state $\alpha\in\mbs$ determines a unique probability measure $m_\alpha:\mbl\to[0,1]$ defined by $m_\alpha(f)=f(\alpha)$, meaning, in particular, that for any pairwise
orthogonal sequence $(f_i)$ in $\mbl$, $m_\alpha(\lor_if_i)
=\sum_im_\alpha(f_i)$. Again, we identify $m_\alpha$ with $\alpha$.

The family of \ $\mbl$-valued measures $\sfm_\sfe$,
$\sfe\in\wmbo$, is {\em surjective} (that is, any $f\in\mbl$ is of the form $f=\sfm_\sfe(X)$ for some $\sfm_\sfe(X)$), 
and the family of probability measures $m_\alpha,\alpha\in\mbs,$ is {\em order determining}, that is,
for any $f,g\in\mbl$, $f\leq g$ if and only if $m_\alpha(f)\leq m_\alpha(g)$ for all $\alpha\in\mbs$.  
For each $\alpha\in\mbs,\sfe\in\wmbo,  X\in\hA$ we have
$$
p(\alpha,\sfe,X)=m_\alpha(\sfm_\sfe(X)).
$$

We note that also the converse result is true: If $\mbl$ is an arbitrary orthocomplemented partially ordered set admitting an order determining set of
probability measure $\mbs$, and $\wmbo$ is a surjective set of $\mbl$-valued measures, then the function $p$ defined as $p(\alpha,\sfm,X)=\alpha(\sfm(X))$ for all
$\alpha\in\mbs, \sfm\in\wmbo, X\in\hA$, is a probability function satisfying the orthogonality postulate and the logic of p is isomorphic to $\mbl$ \cite{Maczynski1973}.

The sets $\mbl_0$ and $\mbl$ of experimental propositions and functions are in one-to-one onto correspondence and 
one may immediately transform the order and complementation of $\mbl$ to  $\mbl_0$: for any $a,b\in\mbl_0$, $a\leq b$ if and only if $f_a\leq f_b$, 
and  we let $a^\perp$ stand for the proposition corresponding to the function $e-f_a$. 
Thus, under the assumption of Axiom~\ref{axiom1}, we may equally well
consider $\mbl_0$ as an
orthocomplemented $\sigma$-orthocomplete orthomodular partially ordered set (of propositions), with $\mbs$ as an order determining set of
probability measures of $\mbl_0$. From now on we do not distinguish between $\mbl_0$ and $\mbl$ and we also interchangeably consider the elements of $\mbl$
as functions on $\mbs$ and the elements of $\mbs$ as functions on $\mbl$: $a(\alpha)=\alpha(a)$. 
Also, together with $\mbl$ we always mean the  structure
$(\mbl,\leq,{}^\perp)$, with the order bounds $0$ and $e$, corresponding to the absurd (always false) and trivial (always true) propositions. 
Moreover, we view the observables $(\sfe,\Omega,\hA)\in\wmbo$  as $\mbl$-valued measures  and we recall that for each $a\in\mbl$ there is an observable $\sfe\in\wmbo$
and a set $X\in\hA$ such that $a=\sfe(X)$.

An important technical assumption concerning the structure of $\mbl$ is the {\em separability of
$\mbl$}; this is the property that any pairwise orthogonal sequence $(a_i)\in\mbl$ is at most countably infinite. 
This structure has the following measurement theoretical justification.
The range $\sfe(\hA)$ of any observable $\sfe\in\wmbo$  is a Boolean sub-$\sigma$-algebra of $\mbl$. 
If the value space $(\Omega,\hA)$ of $\sfe\in\wmbo$ is a subspace of the real Borel space $(\R^n,\hB(\R^n))$, for some $n=1,2,\ldots$, 
then the Boolean $\sigma$-algebra $\sfe(\hA)$ is separable.
By the classic Loomis-Sikorski theorem, any separable Boolean sub-$\sigma$-algebra $\mathcal B\subset\mbl$ is the range of some
(real valued) observable $\sfe:\br\to\mbl$  \cite{Varadarajan1985}. 
If the logic $\mbl$ is separable
then any
Boolean sub-$\sigma$-algebra of $\mbl$ is also separable and thus appears as the range of an observable. 
With this motivation we pose the following assumption:

\begin{itemize}
\item[] {\em Separability of the logic}: Any  orthogonal sequence $(a_i)\subset\mbl$ 
is at most countably infinite.
\end{itemize}

We call the pair $(\mbs,\mbl)$ the {\em logic} of the statistical duality $(\mbs,\mbo,p)$ associated with a subset $\wmbo$ of observables satisfying the orthogonality postulate, Axiom~\ref{axiom1}. 
We also assume that the logic is separable.

\subsection{Further specifications}\label{sect:furtherspecifications}
The set $\mbs$ of states is convex 
and it determines the order on $\mbl$.
We now assume that pure states exist. Moreover, we assume that there are sufficiently many so that each 
 $a\in\mbl, a\ne 0$, can be realized in some pure state, that is, there is an  $\alpha\in\ps$ such that $\alpha(a)=1$.

\begin{itemize}
\item[] {\em Sufficiency of pure states}:
   For any $a\in\mbl$, $a\ne 0$, there is an $\alpha\in\ps$ such that $\alpha(a)=1$.
\end{itemize}

The Jauch-Piron property\footnote{This property has independently been introduced in \cite{Zierler1961} and \cite{JauchPiron1963} and it is known to be equivalent to the fact that each $\alpha\in\mbs$ has  a (unique) support  in $\mbl$ \cite[Theorem 11.4.3]{EnricoGianni}.} 
is a further important property of the logic $(\mbs,\mbl)$:

\begin{itemize}
\item[] {\em The Jauch-Piron property}:  For any $a,b\in\mbl$, if $\alpha(a)=\alpha(b)=1$ for some $\alpha\in\mbs$, then there exists a $c\in\mbl$ such that $c\leq a$, $c\leq b$ and $\alpha(c)=1$.
\end{itemize}

These two assumptions have
strong structural implications. To state the relevant result we recall that an element $a\in\mbl$ is the {\em support}
of the state $\alpha\in\mbs$ if $\alpha(a)=1$ and 
for any $b\in\mbl$ the condition $\alpha(b)=1$ implies $b\geq a$, that is, $a$, if exists, is the smallest proposition which is true (in the sense of probabilistic certainty) in the state $\alpha$. If the support of $\alpha$ exists it is unique and we donote it by $s(\alpha)$.

\begin{theorem}\label{BChila}
If the set $\mbs$ of states of the  logic $(\mbs,\mbl)$ 
contains a sufficient set $\ps$  of pure states and satisfies the Jauch-Piron property then
$\mbl$ is a complete
orthocomplemented orthomodular lattice. Each state $\alpha\in\mbs$ has a support $s(\alpha)\in\mbl$
and each $a\in\mbl, a\ne 0$, is a support of some state $\alpha\in\mbs$. 
\end{theorem}


\noindent
{  {\bf Proof.\footnote{Our proof is an adaption of the corresponding results 
in \cite{EnricoGianni}. Another source leading to this conclusion is given by
the results of Section 2.5.2 
of \cite{PavelSylvia}.}}
We show first that each $\alpha\in\mbs$ has a support in $\mbl$. If $\alpha(a)\ne 0$ for any $a\in\mbl$, then $\alpha(a)<1$ for each $a\ne 1$, meaning that $s(\alpha)=1$.
If $\{a\in\mbl\,|\, \alpha(a)=0\}\ne \{0\}$, we choose by Zorn's lemma a maximal orthogonal family in this set. By the separability of $\mbl$ this family is at most countably infinite. Hence there is a maximal orthogonal sequence $(a_i)_{i\geq 1}$ with $\alpha(a_i)=0$ for all $i$. 
Let $a=\lor_ia_i$ and observe that $\alpha(a)=0$. To establish that  $a^\perp=s(\alpha)$,  we show that for any $x\in\mbl$, $\alpha(x)=0$ if an only if $x\perp a^\perp$, that is, $x\leq a$.
If $x\leq a$, then $\alpha(x)\leq\alpha(a)=0$. To show the converse, assume that $\alpha(x)=0$. By the (dual) Jauch-Piron property there is a $c\in\mbl$ such that $x\leq c, a\leq c$ and $\alpha(c)=0$.
If $x\not\leq a$ then $c\ne a$ (since otherwise $a=c\geq  x$) and thus, by orthomodularity $c=a\lor(c\land a^\perp)$. Therefore, $\alpha(c)=\alpha(a)+\alpha(c\land a^\perp)$ and thus $\alpha(c\land a^\perp)=0$. Since $c\land a^\perp$ is orthogonal to each $a_i$ we may expand the maximal orthogonal sequence $(a_i)$, which is a contradiction. Hence, $x\leq a$,  showing that $a^\perp=s(\alpha)$.

We show next that each $a\in\mbl$, $a\ne 0$, is the support of some $\alpha\in\mbs$. Let $W(a)=\{x\in\mbl\,|\, x=s(\alpha)\ {\rm for\ some}\ \alpha \ {\rm such \ that}\  \alpha(a)=1\}$. By the sufficiency of $\ps$ this is a
nonempty set. Moreover, if $x\in W(a), x=s(\alpha)$ and $\alpha(a)=1$ then $x\leq a$. Let $(x_i)$ be a maximal (countable) orthogonal sequence in $W(a)$ and define $b=\lor_i x_i$ (so that $b\leq a$).
As above, if $b\ne a$, then $a\land b^\perp$ would be an element in $W(a)$ pairwise orthogonal with each $x_i$, which is not possible. Thus $\lor_ix_i=a$. Since any $x_i$ is the support of some $\alpha_i$ (for which $\alpha_i(a)=1$), then $a$ is the support of all the convex combinations $\sum w_ix_i$ (with all $w_i>0$).  

It remains to be shown that $\mbl$ is a complete lattice.  Let $a,b\in\mbl$, $a\ne 0\ne b$ (if $a$ or $b$ is 0 the supremum and
infimum exist trivially). Let $\alpha,\beta\in\mbs$ be such that $s(\alpha)=a, s(\beta)=b$, and consider the state
$\gamma=\lambda\alpha+(1-\lambda)\beta$ for some $0\ne\lambda\ne 1$. Clearly $s(\gamma)=a\lor b$. By De Morgan laws one gets the dual result. 
It is well-known that every separable orthomodular $\sigma$-orthocomplete lattice is complete, see e.g. \cite[Lemma 2.5.2f]{PavelSylvia}. \qed

\

There are three further important properties  the logic $(\mbs,\mbl)$ must possess in order to provide a geometric representation of the elements of $\mbl$
as subspaces of a vector space. The first property is the atomicity: $\mbl$ is {\em atomic} if every $a\in\mbl$, $a\ne 0$, contains an atom. We recall that an element $p\in\mbl$ is an atom
if for any $a\in\mbl, a\ne 0$, the condition $a\leq p$ implies $a=p$. We let $\atl$ denote the set of atoms in $\mbl$.

To get the atomicity of $\mbl$ we pose the following assumption concerning the identification of pure states. In Section~\ref{sect:filters}
this assumption is formulated in terms of operations and it forms  a part of the projection postulate.

\begin{itemize}
\item[] {\em Identification of pure states}: Let $\alpha\in\ps$. For any $\beta\in\mbs$, if $\beta(s(\alpha))=1$ then $\beta=\alpha$.
\end{itemize}

\begin{proposition}\label{purestatesandatoms}
With the assumptions of Theorem~\ref{BChila}, the identification of pure states implies that the support of any pure state is an atom. Moreover, $\mbl$ is atomic and the map
$\ps\ni\alpha\mapsto s(\alpha)\in\atl$ is a bijection.
\end{proposition}

\noindent
{\bf Proof.} Let $p=s(\alpha)$ be the support of $\alpha\in\ps$ and let $a\in\mbl, a\ne 0,$ be such that $a\leq p$. Since $a\ne 0$ there is a $\beta\in\ps$ such that $\beta(a)=1$.
From $a\leq p$ one then gets $\beta(p)=1$, which means that $\beta=\alpha$. Since  $a\geq s(\beta)$  and $s(\beta)=s(\alpha)=p$, one has $a\geq p$ and thus $a=p$, that is, $p$ is an atom.

For $a\in\mbl, a\ne 0$, there is a $\alpha\in\ps$ such that $\alpha(a)=1$. Therefore $s(\alpha)\leq a$, showing that $\mbl$ is atomic.

We leave it as an exercise to show that
the mapping $\ps\ni\alpha\to s(\alpha)\in\atl$ is injective and surjective.\qed

\

The second ingredient required to establish the
vector space realization is the {\em covering property}: for any $a\in\mbl$ and $p\in\atl$, if $a\land p=0$, then $a\lor p$ covers $a$, that is, for any $b\in\mbl$,
if $a\leq b\leq a\lor p$, then $b=a$ or $b=a\lor p$. Since $\mbl$ is an atomic lattice the covering property can equivalently be  formalized as follows:
for any $a\in\mbl,p\in\atl$, the element $(a\lor p)\land a^\perp$ is either an atom or $0$ \cite[Prop. 3.2.17]{PavelSylvia}.
To obtain the covering property for $(\mbs,\mbl)$ we adapt a part of the projection postulate  reflecting the possibility of actualizing potential properties with {\em minimal disturbance}. In Section~\ref{sect:filters} we present a full formulation of this postulate together with an elucidation of its physical motivation. 
The ideality assumption (I1) of a filter given there corresponds to the following minimal disturbance requirement.

\begin{itemize}
\item[] {\em Minimal disturbance}: If $\alpha\in\ps$, $a\in\mbl$, and $\alpha(a)\ne 0$, then there exists a pure state $\beta\in\ps$ such that 
$\beta(a)=1$, that is, $s(\beta)\leq a$, and $\alpha(s(\beta))=\alpha(a)$.
\end{itemize}

\begin{proposition}\label{BB1973thm} (Bugajska, Bugajski, \cite{BB1973})
With the assumptions of Theorem~\ref{BChila} and the identification of the pure states, the minimal disturbance implies the covering property.
\end{proposition}

\noindent
{ { {\bf Proof.}
Let $p\in{\rm At}(\mbl),a\in\mbl$ and $p=s(\alpha)$. Let $\alpha_1$ and $\alpha_2$ be the pure states  
such that $\alpha(a)=\alpha(s(\alpha_1))$ and $\alpha(a^\perp)=\alpha(s(\alpha_2))$ as
given by the minimal disturbance. Clearly, $\alpha(s(\alpha_1)\lor s(\alpha_2))=1$, so that $p\leq s(\alpha_1)\lor s(\alpha_2)$ and $p\lor a^\perp\leq s(\alpha_1)\lor a^\perp$.
Hence $(p\lor a^\perp)\land a\leq (s(\alpha_1)\lor a^\perp)\land a=s(\alpha_1)$, which means that  $(p\lor a^\perp)\land a$ is either the atom $s(\alpha_1)$ or $0$, that is, the covering property holds in $\mbl$.  \qed 
}}

\

We collect the above assumptions concerning the set of states of the logic  in the form of an axiom.

\begin{axiom}\label{axiom2} 
The set $\mbs$ of states of the logic $(\mbs,\mbl)$,
with a separable $\mbl$,
has a sufficient set of pure states, the Jauch-Piron property, and it allows the identification of pure states and the minimal disturbance.  
\end{axiom}

An atomic lattice with covering property is often referred to as an AC lattice.
We may thus conclude that
Axioms~\ref{axiom1} and \ref{axiom2} 
imply that $\mbl$ is a complete orthomodular AC lattice and that the support function gives a bijective correspondence between the sets $\ps$ and ${\rm At}(\mbl)$.

The final ingredient required to establish a geometric representation of  $(\mbs,\mbl)$  is the irreducibility
 of $\mbl$.

\subsection{The classical case excluded}\label{sect:classicalexcluded}

There are various features of quantum mechanicsthat have been elevated   to the status of   fundamental principles
of the theory. 
These include the notions of superposition, complementarity, uncertainty,  entanglement, 
nonunique decomposability and purification of mixed states, 
and irreducibility of  probabilities, which   are unquestionably among the most widely discussed 
characteristic traits of quantum mechanics. 
In Remark~\ref{classicalcase} the unique decomposability of mixed states into its pure components is seen to be closely related to a classical description and it goes  hand in hand with the Boolean structure of the set of decision effects.
Here we discuss briefly the notions of superposition and complementarity to show that for a proper quantum system the logic $\mbl$ is far from being Boolean.
We  can draw on  the full structure of the pair $(\mbs,\mbl)$ introduced in the preceding subsections~\ref{OMsection} and \ref{sect:furtherspecifications}
even though not all of it is actually needed here.

\subsubsection*{Superpositions}
There are several formulations of the notion of superposition in  quantum logic.
We adopt the following definition
taken from \cite[p. 53]{Varadarajan1985} 
as a formalization of the intuitive ideas of Dirac \cite{Dirac1958}:

\begin{definition}\label{SP1}
A pure state $\alpha\in\ps$ is a \emph{superposition} of pure states $\alpha_1,\alpha_2\in\ps$ if and only if $\alpha_1(a)=\alpha_2(a)=1$ implies $\alpha(a)=1$ for every $a\in\mbl$.  Equivalently, a pure state $\alpha$ is a superposition of pure states $\alpha_1$ and $\alpha_2$ if and only if $s(\alpha)\leq s(\alpha_1)\lor s(\alpha_2)$.
\end{definition}

Instead of stating directly a superposition principle 
we give the following definition:
\begin{itemize}
\item[]
A physical system with the structure $(\mbs,\mbl)$ is a {\em proper quantum system} if 
for every two pure  states $\alpha,\beta\in\ps$, $\alpha\ne\beta$, there exists a third one $\gamma\in\ps$, $\alpha\ne\gamma\ne\beta$, which is their superposition. 
\end{itemize}
It is then a simple, but important consequence that the logic of
a proper quantum system is {\em irreducible}, 
that is, the centre of $\mbl$
$$
{\rm Cent}(\mbl) =\{c\in\mbl\,|\ a=(a\land c)\lor (a\land c^\perp)\ {\rm for\ any}\ a\in\mbl\}
$$
contains only the trivial elements $0$ and $e$. (For a proof, see e.g. \cite[Cor. 3.2.4]{PavelSylvia}.)

\begin{remark}\rm
Note that  if $c\in{\rm Cent}(\mbl)$, then for any $a\in\mbl$, there is
an observable $\sfe\in\wmbo$ such that $c=\sfe(X)$ and $a=\sfe(Y)$ for some value sets $X$ and $Y$. 
Thus for a proper quantum system
there is no (nontrivial)
proposition (or property) that could be measured together with evry other proposition (or property).
By contrast, if ${\rm Cent}(\mbl)=\mbl$, the set $\mbl$ forms a Boolean $\sigma$-algebra for which the theorems of Stone \cite{Stone} and Loomis \cite{Loomis} and Sikorski \cite{Sikorski}
give a representation as a $\sigma$-algebra $\hA$ of subsets of a set $\Omega$.
\end{remark}

\begin{remark}\rm
If a pure state $\alpha$ is a superposition of pure states $\beta$ and $\gamma$, then also $\beta$ is a superposition of $\alpha$ and $\gamma$,
and likewise $\gamma$ is a superposition of $\alpha$ and $\beta$. This is the exchange property and it is often included in the notion of superposition of states. In the present
context this property is equivalent to the covering property, obtained above from the projection postulate. For a proof, see e.g. \cite[Prop. 3.2.17]{PavelSylvia}.
\end{remark}

\subsubsection*{Complementarity}
The existence of pairs of complementary observables is another fundamental feature of quantum mechanics. Following the ideas of Bohr \cite{Bohr1935}
we say that two observables are complementary if all the experimental arrangements which unambiguously define these observables are mutually exclusive. Again, 
there are various ways of formalizing this intuitive idea.
We adopt the following definition appropriate to the logic $(\mbs,\mbl)$.

\begin{definition}\label{CP1}
Properties $a,b\in\mbl$ are {\em complementary} if they are disjoint, that is, $a\land b =0$, but not orthogonal, that is, $a\not\leq b^\perp$.
Equivalently, $a,b\in\mbl$ are {\em complementary}, if for any $\alpha\in\mbs$, the condition $\alpha(a)=1$ implies $0\ne\alpha(b)\ne 1$, and the condition $\alpha(b)=1$ implies $0\ne \alpha(a)\ne1$.  
\end{definition}
In a Boolean logic, the conditions $a\land b=0$ and $a\leq b^\perp$ are equivalent. This means that if there are complementary properties 
in $\mbl$ then $\mbl$ cannot be Boolean.

It is a simple exercise to show that $\mbl$ is irreducible if
 for any $a\in\mbl$, $a\ne 0,e$,  there is a $b\in\mbl$ such that $a$ and $b$ are complementary.
As an alternative to the previous definition, we could call
 a physical system with the structure $(\mbs,\mbl)$ a {\em proper quantum system} if 
for any $a\in\mbl$, $a\ne 0,e$, there is a $b\in\mbl$ such that $a$ and $b$ are complementary.
It then follows that for a proper quantum system  the logic $\mbl$ is irreducible.

We summarize the main result of this section.

\begin{corollary}\label{AClattice1} Let $(\mbs,\mbo,p)$ be the statistical duality
satisfying Axioms~\ref{set_of_states} and \ref{affine}. If 
the logic $(\mbs,\mbl)$ defined by a (nonempty) subset $\wmbo$ of observables  satisfies Axioms~\ref{axiom1} and \ref{axiom2} and
 the physical system in question is a proper quantum system, then $\mbl$ is an irreducible complete orthomodular AC lattice
and there is a bijective correspondence (given by the support function) between the set $\ps$   of
pure states in $\mbs$  and the set  ${\rm At}(\mbl)$ of atoms in $\mbl$.
\end{corollary}

\section{Filters and the projection postulate}\label{sect:filters}

Assume now that the statistical duality satifies the stronger Axioms~\ref{statespace} and \ref{effects}.
The existence of a sufficiently large subset of observables $\wmbo\subset\mbo$ leading to the fundamental result of Corollary~\ref{AClattice1}
goes together with the existence of a subset $\widetilde{\bm{\hO}}$ of operations such that $e\circ\widetilde{\bm{\hO}}=\mbl$. This suggests that the same conclusion could be reached by singling out a sufficiently large and regular set of operations. 
This is what we consider next. 

\begin{remark}\label{ifkm}\rm  With the structure specified by Theorem~\ref{BChila} and Propositions~\ref{purestatesandatoms} and \ref{BB1973thm} one may construct for each $a\in\mbl$ a map $\phi_a:\mbs\to\mbs$ whose restriction on pure states
is uniquely defined by $s(\phi_a(\alpha))=(s(\alpha)\lor a^\perp)\land a$, $\alpha\in\ps$, and which has the typical properties of a state transformation caused by an ideal first kind measurement
\cite{Pool1968,Pool1968b,CassinelliBeltrametti1975}. Due to the properties of the support function $s:\mbs\to\mbl$ the map $\phi_a$ fails to be linear, 
that is,  it is not an operation in the sense of Section~\ref{sect:operationseffects}. Apart from this the characteristic properties of such a $\phi_a$ serve below as the defining properties of filters. 
\end{remark}


\subsection{Filters}
Filters are a special kind of operations reflecting certain ideality properties that the so-called yes-no (or {\em simple}) measurements may or may not possess.
The properties of filters have  been discussed extensively in the literature, see, e.g. \cite{Pool1968,Pool1968b,DaviesLewis1970,Davies1976,Edwards1971,Edwards1972,Mielnik1974,CassinelliBeltrametti1975,EnricoGianni,BugLah1985}. 
This allows us to be brief in their introduction.  The properties of filters are defined through their action on pure states. 
The definition thus presumes that the set $\ps={\rm ex}(\mbs)$ of pure states is not empty.

An operation $\phi\in\bm\hO$ is  {\em pure} if 
\begin{itemize}
\item[(P1)]\quad  $\phi(\alpha)\in[0,1]\cdot\ps$  for any pure state $\alpha\in\ps$,
\end{itemize}
and
an operation $\phi$ is of the {\em first kind} if
\begin{itemize}
\item[(F1)]\quad  $e(\phi(\alpha))=1$ implies $\phi(\alpha)=\alpha$ for any $\alpha\in\ps$,
\item[(F2)]\quad $e(\phi^2(\alpha))=e(\phi(\alpha))$ for any $\alpha\in\ps$.
\end{itemize}
To define the ideality of an operation we
first assume that any pure state can be identified by an operation:
\begin{itemize}
\item[(S1)] \quad  for any pure state $\alpha\in\ps$ there is a unique  $\phi_\alpha\in\bm{\hO}$ such that $e(\phi_\alpha(\beta))=1$ implies $\beta=\alpha$
for any $\beta\in\ps$.
\end{itemize}
We then say that  a pure operation $\phi$ is {\em ideal} if
\begin{itemize}
\item[(I1)]\quad $e(\phi(\alpha))=e(\phi_{\alpha'}(\alpha))$ for any $\alpha\in\ps$, 
with $\phi(\alpha)\ne0$, where  $\alpha'=e(\phi(\alpha))^{-1}\phi(\alpha)$ and $\phi_{\alpha'}$ as in (S1).  
\end{itemize}

A  pure, ideal, first kind operation is a {\em  filter} and
we let $\bm{\hO_f}$ denote the set of filters.
We comment briefly on the defining properties of filters. 

The purity (P1) of an operation means that it takes a pure state to a pure state with a possible loss in the intensity. As pure states may be interpreted as maximal information states,
a pure operation leaves the system in a maximal information state whenever it was in such a state. 

With the so-called ideality assumptions one usually aims at minimizing the influence on the state caused by an operation performed on the system. In addition to the purity condition (P1)
and the first kind conditions (F1) and (F2), the condition (I1) aims at that. It claims that an ideal $\phi$ maps any pure state $\alpha$ onto an eigenstate of $\phi$  closest to $\alpha$,  thus  disturbing the system to a minimal extent. This is  the minimal disturbance assumption of Section~\ref{sect:furtherspecifications}.

Of the two first kind conditions (F1) and (F2), (F1) claims that if $\phi$ does not lead to a detectable effect when performed on the system in a pure state $\alpha$ then, provided that the
operation is good enough, it does not alter the state of the system, either. According to (F2), a repeated application of a good operation does not lead to a new effect.


As an immediate consequence of the defining properties of filters, we note that they are not only weakly repeatable ($e(\phi^2(\alpha))=e(\phi(\alpha))$ for any $\alpha\in\ps$) but
also repeatable ($\phi^2(\alpha)=\phi(\alpha)$ for any $\alpha\in\ps$). 
 Moreover, filters satisfy the most common ideality requirement: if a good operation $\phi_1$ is performed on the system in a pure state $\alpha$ which is an eigenstate of a good operation $\phi_2$ (i.e. $e(\phi_2(\alpha))=e(\alpha)$) which commutes weakly with $\phi_1$ (i.e. $\phi_1\circ\phi_2$ and $\phi_2\circ\phi_1$ are isotonic), then $\phi_1$ leaves the system in a state which is still an eigenstate of $\phi_2$.

We say that the set $\bm{\hO_f}$ of filters is {\em sufficiently rich} if the operations of (S1) are filters
and 
\begin{itemize}
\item[(S2)]\quad  for each filter $\phi\in\bm{\hO_f}$ there is another filter $\phi'\in\bm{\hO_f}$ such that  $e\circ\phi' =(e\circ\phi)^\perp$.
\end{itemize}
Condition (S1), the identification of pure states, expresses the common belief that any pure state $\alpha$ can be produced by a particular selection or filtering process $\phi_\alpha$, 
which under the conditions (F1) and (F2) receives the form $\phi_\alpha(\beta)=e(\phi_\alpha(\beta))\alpha$ for any $\beta\in\ps$. 
The second sufficiency condition (S2) stipulates that
%
if an effect $a$ can be obtained from a filter, that is $a=e\circ \phi$ for some $\phi\in\bm{\hO_f}$, then also its `negation' $a^\perp=e-a$ 
can  be produced by a pure ideal first kind operation.

\subsection{Projection postulate}
The set $\mbl$ of {\em propositions} of the convex scheme $(\mbs,\bm{\hO},\mbe)$ is now defined as the set of all decision effects $a\in\pmbe$ with nonempty certainly-yes-domain $a^1=\{\alpha\in\ps\,|\, a(\alpha)=1\}$ together with the null effect $0$,
$$
\mbl =\{a\in\pmbe\,|\, a=0\ {\rm or}\ a^1\ne\emptyset\}.
$$

For a given system $(\mbs,\bm{\hO},\mbe)$ the set $\bm{\hO_f}$ of filters may be empty and the set $\mbl$ of propositions may be trivial $\{0,e\}$. However, for any 
$\phi\in\bm{\hO_f}$, $\phi\ne 0$, the resulting effect $e\circ\phi$ has a nonempty certainly-yes-domain $(e\circ\phi)^1$. 
By Remark~\ref{ifkm} it is also natural to expect that for any $a\in\mbl$, $a\ne 0$, there is a  filter $\phi_a$ such that $e\circ\phi_a=a$.
With the projection postulate we confirm this expectation together with a unicity assumption.   
In that we also assume that the set of pure states is not only nonempty but is also {\em strongly ordering} on $\mbl$, that is, it is ordering and for any $f,g\in\mbl$, if $f^1\ne\emptyset$, and $f^1\subseteq g^1$, then $f\leq g$.

\begin{axiom}\label{PP}
The statistical duality $(\mbs,\mbo,p)$ of Axioms \ref{statespace} and \ref{effects} 
satisfies the {\em projection postulate} if
the set $\ps$ of pure states is strongly ordering on $\mbl$, the subset of filters $\bm{\hO_f}\subset\bm{\hO}$ is sufficiently rich and 
there is a bijective mapping  $\Phi:\mbl\to\bm{\hO_f}$ with the property:
$a(\alpha)=e(\Phi(a)(\alpha))$ for every $a\in\mbl$ and $\alpha\in\ps$.
\end{axiom}

The projection postulate guarantees the existence of a sufficiently rich collection of the important class of operations associated with the pure, ideal, first-kind measurements, but it does not restrict the theory to  such measurements only. 
Neither does it distinguish between classical and quantum descriptions. In any case,
this postulate has strong structural implications on the order structure of the set $\mbl$ of propositions. They will be studied next.

\begin{lemma}
For a statistical duality $(\mbs,\mbo,p)$ satisfying the projection postulate,
the set $\mbl$ of propositions is a nonempty partially ordered set with $a\mapsto a^\perp$ as orthocomplementation.
\end{lemma}

\noindent
{\bf Proof.}
Since $\ps\ne\emptyset$ the set $\bm{\hO_f}$ of filters is  nonempty and thus also $\mbl\ne\emptyset$. The set ${\rm ex}(\mbe)$ of
decision effects is closed under the map $a\mapsto a^\perp=e-a$. If $a\in\mbl$, 
with $a=e\circ\Phi(a), \Phi(a)\in\bm{\hO_f}$,
then by (S2) $a^\perp=e\circ\Phi(a)'$ for some
$\Phi(a)'\in\bm{\hO_f}$, so that $a^\perp\in\mbl$. Clearly, $\Phi(a)'=\Phi(a^\perp)$. Let $b\in\mbl$ be such that $b\leq a$ and $b\leq a^\perp$, and
assume that $b\ne 0$. Then for any $\alpha\in b^1$, $a(\alpha)=1$ and $a^\perp(\alpha)=1$ which is impossible. Thus $b=0$, that is, $a\land_{\mbl}a^\perp =0$.
The remaining claims are obvious. \qed

\  

\begin{lemma}
For any $a,b\in\mbl$, if $a\leq b^\perp$, then $a+b\in\mbl$. 
Similarly, for any triple $(a,b,c)$  of mutually orthogonal elements of $\mbl$, $a+b,a+c,b+c,a+b+c\in\mbl$.
\end{lemma}
\noindent
{\bf Proof.}
For $a\perp b$, $a+b\leq e$. Now $(a+b)^1\supseteq a^1\cup\ b^1\ne\emptyset$. Assume that $a+b=\frac 12(f+g)$ for some $f,g\in\mbe$.  
Then $(a+b)^1=f^1\cap g^1$ so that $a+b\leq f$ and $a+b\leq g$. Thus $a+b=a+b+\frac 12(f-(a+b) + g -(a+b)),$ which implies that $f=g=a+b$, that is, $a+b\in{\rm ex}(\mbe)$. Hence
$a+b\in\mbl$. Copying the argument for a triple of mutually orthogonal elements $a,b,c\in\mbl$ one immediately concludes also that $a+b+c\in\mbl$.
\qed

\ 

\begin{corollary}\label{Cor13082012}
For a statistical duality $(\mbs,\mbo,p)$ satisfying the projection postulate, the set 
$\mbl$ of propositions is orthomodular.
\end{corollary}
\noindent
{\bf Proof.}
Let $a,b,c\in\mbl$ be a triple of mutually orthogonal elements. Then not only $a+b,a+c$ and $b+c$  but also $a+b+c\in\mbl$. This means that $\mbl$ is triangle-closed in the sense of \cite{LahtiMaczynski1992}. By \cite[Theorem 3.2]{LahtiMaczynski1992} this is equivalent to $\mbl\subseteq\mbe$ being orthomodular, and, in particular, $a+b=a\lor_{\mbl}b$ for $a,b\in\mbl, a\leq b^\perp$. \qed

\ 

\begin{lemma}
For any $\alpha\in\ps$, $e\circ\phi_\alpha\in{\rm At}(\mbl)$. Moreover, $\mbl$ is atomic, that is, any $a\in\mbl$, $a\ne 0$,  contains an atom.
The map $\ps\ni\alpha\mapsto e\circ\phi_\alpha\in{\rm At}(\mbl)$ is a bijection, with $e\circ\phi_\alpha$ being the support of $\alpha$.
\end{lemma}
\noindent {\bf Proof.}
Let $a\in\mbl$, $\alpha\in\ps$, and assume that $a\leq \Phi^{-1}(\phi_\alpha)$. If $a\ne 0$ then for any $\beta\in a^1$, $\Phi^{-1}(\phi_\alpha)(\beta)=e(\phi_\alpha(\beta))=1$, so that by (S1) $\beta=\alpha$, that is $a^1=\{\alpha\}$. Therefore $\Phi(a)=\phi_\alpha$, or equivalently,
$a=\Phi^{-1}(\phi_\alpha)$, which entails that for any $\alpha\in\ps$, $\Phi^{-1}(\phi_\alpha)$ is an atom. Clearly, for any $a\in\mbl$, one has
$\Phi^{-1}(\phi_\alpha) =e\circ\phi_\alpha\leq a$ for all $\alpha\in a^1$. \qed

\

\begin{lemma}
The set $\ps$ of pure states is sufficient for $\mbl$, and $\mbl$ has the Jauch-Piron property.
\end{lemma}
\noindent {\bf Proof.}
The sufficiency of $\ps$ for $\mbl$ is obvious. Let $a,b\in\mbl$ be such that $a^1\cap b^1\ne\emptyset$. For any $\alpha\in a^1\cap b^1$, $e\circ\phi_\alpha$ is contained both in $a$ and in $b$ and $e(\phi_\alpha\alpha)=1$. \qed

\  

We observe that the range $\sfe(\hA)$ of an observable $\sfe\in\mbo$ is Boolean if it is contained in $\mbl$. Therefore we may again justify {\em the separability assumption} of $\mbl$ with the requirement that any Boolean subsystem of $\mbl$ could be realized as the range of an  observable. With the separability assumption, $\mbl$ thus acquires the structure specified in Theorem~\ref{BChila}. 
By virtue of Proposition~\ref{BB1973thm}, the  ideality property (I1) of filters  then gives the covering property. Hence we have the following.

\begin{theorem}
If the operational description $(\mbs,\bm{\hO},\mbe)$  defined by Axioms \ref{statespace} and  \ref{effects} satifies the projection  postulate 
and $\mbl=\{a\in{\rm ex}(\mbe)\,|\, a^1\ne\emptyset\  {\rm or}\  a=0\}$ is separable, 
then $\mbl$ forms a complete atomic orthomodular orthocomplemented lattice with the covering property. Moreover, the support function gives a bijective
correspondence between the pure states in $\mbs$ and the atoms of $\mbl$.
\end{theorem}

 To get the irreducibility of $\mbl$ it is most straightforward to require that any two pure states can be superposed into a new pure state. With the structures given by the projection postulate we may immediately adopt Definition~\ref{SP1} to conclude that for a proper quantum system 
the structures of Corollary~\ref{AClattice1} are again available. 

It is to be emphasized, however, that even though the two sets of axioms [\ref{set_of_states},\ref{affine},\ref{axiom1},\ref{axiom2}] and [\ref{statespace},\ref{effects},\ref{PP}] lead to the common structure of Corollary~\ref{AClattice1}, the first approach starts with the weaker assumptions concerning the pair $(\mbs,\mbe)$. Therefore, it is concievable that there are pairs $(\mbs,\mbl)$ with the structure of this corollary appearing as models for the first approach but not for the second approach.

\section{Hilbert space coordinatization}\label{HScoordinatization}

\subsection{The basic problem}
In the Hilbert space formulation of quantum mechanics the pair
$(\mbs,\mbe)$  is given as
the sets of density operators and effect operators on  a complex separable Hilbert space, whereas
$\mbl$ is identified as the set of  (orthogonal) projections on it. It is a deep theorem of Gleason  \cite{Gleason1957}
which assures that all the probability measures on $\mbl$ arise from the density operators through the familiar trace formula.\footnote{For a detailed discussion of this  theorem, see, e.g. \cite{Varadarajan1985}}

Untill now we have presented two sets of axioms for the structures $(\mbs,\mbe)$ and   $(\mbs,\mbl)$ associated with a proper quantum system. 
The remaining problem of this axiomatic approach is to show that the only realization of this abstract structure is the one given by the Hilbert space quantum mechanics.
In the following we present an outline of the solution of this problem, including some still open, critical points.

The traditional way of approaching the problem has been to isolate first the structure of $\mbl$ 
and to look for the models of this structure  alone.
Then, only after having obtained the models of $\mbl$, the structure of $\mbs$ is added and $\mbe$ is determined.
One might expect that this way of voluntarily neglecting a good part of  the basic structures of the pairs $(\mbs,\mbe)$ and $(\mbs,\mbl)$ cannot be the most optimal approach. We return to this question later.

\subsection{The fundamental representation theorem}\label{ssect:reptheorem}
Let $K$ be   a division ring with an involutive antiautomorphism $\lambda\mapsto\lambda^*$ (such that $(\lambda+\mu)^*=\lambda^*+\mu^*, (\lambda\mu)^*=\mu^*\lambda^*,\lambda^{**}=\lambda$) and let $V$ be a (left) vector space over $K$. A {\em Hermitian form}\index{form!Hermitian} on $V$ is
a mapping $f:V\times V\to K$ with the following properties: for any $u,v,w\in V, \lambda,\mu\in K$,
\begin{eqnarray*}
&&f(\lambda u+\mu v,w)=\lambda f(u,w)+\mu f(v,w)\\
&&f(u,v)^*=f(v,u)\\
&&f(v,v)=0\ {\rm implies}\ v=0.
\end{eqnarray*}
If $V$ admits a Hermitian form $f$ we say that $V$, or rather $(V,K,{}^*,f)$, is a {\em Hermitian space}\index{Hermitian space}. A subspace $M\subset V$ of a Hermitian space
is $f$-{\em closed}\index{Hermitian space!closed subspace} if $M=M^{\perp\perp}$, where 
$$
M^\perp=\{v\in V\,|\, f(v,x)=0\ {\rm for\ all}\ x\in M\}.
$$

Let $\mbl_f(V)$ denote the set of all $f$-closed subspaces of $V$. In addition to the trivial subspaces $\{0\}$ and $V$ any finite dimensional subspace is $f$-closed.
Clearly, if $V$ is infinite dimensional they do not exhaust   the set $\mbl_f(V)$.
The subset inclusion $\subseteq$ together with the map $M\mapsto M^\perp$ give $\mbl_f(V)$ the structure of an irreducible complete orthocomplemented AC lattice.
The converse result is a fundamental representation theorem of projective geometry, proved in detail, for instance, in \cite[Theorem 34.5]{Maeda}:

\begin{theorem}
If $\mbl$ is an irreducible complete orthocomplemented AC lattice of lenght
at least 4, that is, the lenght of a maximal chain is $\geq 4$, then there is a 
Hermitian space $(V,K,{}^*,f)$ such that $\mbl$ is ortho-isomorphic to the lattice $\mbl_f(V)$.
\end{theorem}

\

A Hermitian space $(V,K{},^*,f)$ is {\em orthomodular}\index{Orthomodular space} if for any $M\in\mbl_f(V)$,
$$
M+M^\perp =V.
$$
A Hermitian space $(V,K,{}^*,f)$  is known to be orthomodular if and only if the lattice $\mbl_f(V)$ is orthomodular, see e.g. \cite[Theorem 2.8]{Piziak}. Thus we have the following corollary:

\begin{corollary}\label{basiccorollary} 
Assume that $\mbl$ 
is an irreducible complete orthocomplemented orthomodular AC lattice of   length at least 4.
Then there is an orthomodular space $(V,K,{}^*,f)$ such that $\mbl$ is ortho-isomorphic to $\mbl_f(V)$,
in short, $\mbl\simeq\mbl_f(V)$. 
In particular, all the finite dimensional subspaces of $V$ are in $\mbl_f(V)$ and 
the atoms of $\mbl_f(V)$ are the one-dimensional subspaces of $V$.
\end{corollary}

The pure states $\alpha\in\ps$  are in one-to-one onto correspondence with 
the atoms  $[v] =\{\lambda v\,|\, \lambda\in K\} \in\mbl_f(V)$ 
and they are uniquely determined by their values on the atoms, that is, by the numbers $\alpha_{[v]}([u])\in[0,1]$, $[u]\in {\rm At}(\mbl_f(V))$. 
It is to be stressed that this corollary does not yet give any information on the structure of the real numbers  $\alpha_{[v]}([u])$; in particular, it is not known if $\alpha_{[v]}([u])$ could be related to the $K$-number $f(u',v')$ for some $v'\in[v],u'\in[u]$.
If such a conclusion could be reached then $K$ should be an extension of $\R$.

The well-known models for an orthomodular space $(V,K,{}^*,f)$ are the classical Hilbert spaces $\hi$  over $\R$, $\C$, or $\quat$, the quaternions.
In these models, the form $f$ is the scalar product on $V$ and by Gleason's theorem the probabilities $\alpha_{[v]}([u])$ 
are of the form $\alpha_{[v]}([u])= |f(v',u')|^2$ for any $v'\in[v],u'\in[u]$ with $f(v',v')=f(u',u')=1$, provided that $\dim(\hi)\geq 3$ which is the case in Corollary~\ref{basiccorollary}.
However, the  Hilbert spaces do not exhaust the orthomodular spaces.
In the finite dimensional case this is evident, as shown by a simple example.

\begin{example}\label{Qcase}\rm
 Any finite dimensional Hermitian space $(V,K,{}^*,f)$ is orthomodular and each subspace $M$ of $V$ is $f$-closed, see, e.g. \cite{Holland1995}. Hence the lattice
$\mbl_f(V)$ of $f$-closed subspaces 
coincides with the lattice $\mbl(V)$ of all subspaces of $V$, which is modular (and thus also orthomodular). It is obvious that the space $(V,K,{}^*,f)$ need not be a Hilbert space. 
To witness,  consider the  finite-dimensional rational vector space  $\Q^n$  with the natural form $f(\q,\p)=\sum_{i=1}^nq_ip_i$.  
The form $f$ is Hermitian so that $(\Q^n,\Q,{}^{\rm id},f)$ is an orthomodular space.
Clearly $\Q^n$ is not complete with respect to the distance defined by $f$.
 We return to this example in Sect.~\ref{finitecase} where we study probability measures on $\mbl_f(\Q^n)$.
\end{example}

This example leaves open the infinite dimensional case.  In his seminal paper  \cite{Keller}   Keller was able to construct an explicit example of an infinite dimensional orthomodular space that is very far from being a Hilbert space. Further examples emerged later \cite{GrossKunzi}, and we now know that there are plenty of orthomodular spaces other than the classical Hilbert spaces.  The problem then arises to characterize the Hilbert spaces among the orthomodular spaces. This is solved in the next subsection.

\subsection{A theorem of Sol\' er}\label{ssect:soler}
\begin{theorem}\label{Solertheorem}[Sol\' er, \cite{Soler1995}]
Let $(V,K,{}^*,f)$ be an orthomodular space.
The division ring $K$ is either $\R$, $\C$ or $\quat$ and $(V,K,{}^*,f)$ is the corresponding Hilbert 
space\footnote{If $K=\R$ then ${}^*$ is the identity. For $K=\C$ the map ${}^*$ cannot be the identity and if it is continuous then it is the complex conjugation. For $K=\quat$ the map is the quaternionic conjugation. 
} 
if and only if  there is
an infinite sequence of nonzero vectors $e_i$, $i=1,2,\cdots$ such that $f(e_i,e_j)=0$
for all $i\ne j$,  with the property $f(e_i,e_i)=f(e_j,e_j)$ for all $i,j$.
\end{theorem}
This remarkable result  characterizes  the Hilbert space models of the orthomodular spaces  in a, perhaps, unexpected way.
We emphasize that in this theorem  $V$ is required to be {\em infinite dimensional} and {\em orthomodular}. The next two examples demonstrate that neither of these assumptions can be relaxed.

\begin{example}\rm
The vectors $(1,0,\ldots,0),\ldots (0,\ldots,0,1)$ form an orthonormal basis in $(\Q^n,\Q,{}^{\rm id},f)$  but the space, though orthomodular,  is not a Hilbert space.
\end{example}

\begin{example}\rm
Consider  the infinite dimensional vector space $V=\ell_2(\Q)$ of the square summable sequencies of rational numbers $\q = (q_1,q_2,q_3,\ldots)$ with the Hermitian form 
$f(\mathbf q,\mathbf p)=\sum_{i=1}^\infty q_ip_i$.
The lattice $\mbl_f(V)$ of $f$-closed subspaces is a complete, 
irreducible AC lattice of infinite lenght but it is not orthomodular. The vectors $(1,0,\ldots,0,\ldots)$ \ldots $(0,\ldots,0,1,0,\ldots)$ \ldots form an orthonormal basis in $V$ which is not a Hilbert space.
\end{example}

For $\mbl\simeq\mbl_f(V)$ the existence of a sequence of mutually orthogonal vectors $(e_i)$ in $V$ follows from the assumption that $\mbl$ contains an infinite sequence of pairwise orthogonal atoms.
Such an assumption is physically well motivated e.g. by the spectroscopic data or by the assumption that the quantum system can be localized in an Euclidean space.
It is then worth stressing that, 
contrary to our intuition that comes from using complex numbers, it is
the {\em  `norm' requirement} $f(e_i,e_i)=f(e_j,e_j)$ that is here highly non-trivial.
Indeed, suppose that 
$f(e_i,e_i) = \lambda$ and $f(e_i,e_j)=0$. We have to find an element $\mu\in K$ such that $\mu f(e_j,e_j)\mu^*=\lambda$; in this way $f(\mu e_j,\mu e_j)=f(e_i,e_i)$, see Sect.~\ref{finitecase}.
This is a quadratic equation in $K$ that cannot be solved in general.  In $\R$ or $\C$ one would simply take the square root 
of the positive number $\lambda\lambda^*$ whereas e.g. in $\Q$ this would not work.
For instance, a one-dimensional subspace $[\q]=\{\lambda\q\,|\, \lambda\in\Q\}$ of $(\Q^n,\Q,{}^{\rm id},f)$ 
contains a unit vector only if $\sqrt{\sum q_i^2}$ is rational.

Combining Corollary \ref{basiccorollary} with the theorem of Sol\'er we get the following:

\begin{theorem}\label{Solermodified}
Assume that $\mbl$ is an irreducible complete orthocomplemented orthomodular AC lattice that contains an infinite sequence of orthogonal atoms. Then there is an orthomodular space $(V,K,{}^*,f)$ such that $\mbl$ is orthoisomorphic to $\mbl_f(V)$. $K$ is $\R,\C,$ or $\quat$ and $(V,K,{}^*,f)$ is the corresponding Hilbert space if and only if 
$V$ contains an infinite sequence of mutually orthogonal vectors $(v_i)$ with the property
\begin{equation}\label{Sconditions2}
f(v_i,v_i)=f(v_j,v_j) \quad {\rm for\ all}\ i,j.
\end{equation}
\end{theorem}

By assumption,
there is an infinite sequence of 
orthogonal vectors. 
The  essential question is which properties of $(\mbs,\mbl)$  would imply that such a sequence could be chosen to have  the `norm' property \eqref{Sconditions2}.

 Purely lattice theoretical conditions on $\mbl$ are known that are sufficient to ensure  that $\mbl\simeq\mbl(\hi)$ for a Hilbert space $\hi$.
We can refer to the so-called 
 'angle bisection property' \cite{Morash} or the existence of 'harmonic conjugate' pairs of atoms \cite{Wilbur,Holland1995}. 
 They are of geometric nature and, in the light of the present understanding, they seem to lack any physical interpretation.  Therefore, they are not useful for
 the axiomatic scheme followed here.

The  necessary and sufficient conditions for the conclusion $\mbl\simeq\mbl(\hi)$ of Theorem~\ref{Solermodified} are expressed in terms of $(V,f)$.
One might expect that the assumptions of this theorem together with the full structure of the pair $(\mbs,\mbl)$, in particular, the bijection between
$\ps$ and  ${\rm At}(\mbl)$, could already force $\mbl$  to be a Hilbertian lattice. 
We investigate some aspects of this question in
Sect.~\ref{sec:symmetries}, although to the best of our knowledge, this problem remains still largely open.

 The other two remaining questions are: what can be said if $\mbl$ has only a finite length, and how can the states be represented, once we have represented $\mbl$. 

As concerns the latter question we recall that
if $(V,K,{}^*,f)$ is a classical Hilbert space of dimension at least 3, then
all the probability measures on $\mbl =\mbl_f(V)$ are described by  Gleason's theorem. 
According to it, for any probability measure $\alpha$ on $\mbl_f(V)$, there is a unique positive trace one operator  $\rho:V\to V$ such that, for any 
$M\in\mbl_f(V)$, we have $\alpha(M)=\tr{\rho P_M}$, where $P_M$ is the projection onto $M$. For $\dim(V)=2$, the set of all probability measures on $\mbl_f(V)$ is, however,
 much bigger than those defined by the density operators. But
these additional probability measures are not supported by $\mbl_f(V)$; for details, see \cite[Sect. 25.2]{EnricoGianni}. To the best of our knowledge, there is  no empirical evidence which would require the use of such probability measures as states of a two-level quantum system.

The situation is very different when $(V,K,{}^*,f)$ is not a classical Hilbert space. Very little is known of the probability measures on the lattices $\mbl_f(V)$. 
Keller  \cite{Keller1984} gives examples of nonclassical $\mbl_f(V)$ 
for which one may construct a rich supply of probability measures $m:\mbl_f(V)\to[0,1]$, see also \cite[Problem 7]{GrossKunzi}.
No classification theorem of the Gleason type is available for these examples, and it also
seems that for them there is no one-to-one correspondence between pure probability measures and atoms of $\mbl_f(V)$. 

  \subsection{Finite dimensional case: an example}\label{finitecase} 
 As already noted above, the  structure of a quantum logic $(\mbs,\mbl)$ satisfying $\mbl\simeq\mbl_f(V)$ with $\dim(V)<\infty$   may be substantially different from the infinite dimensional case.
 To emphasize this further we continue Example \ref{Qcase} with determining the set of states $\mbs$ for the logic $\mbl_f(\Q^n)$.

Consider the rational orthomodular space $\Q^n$ with the lattice $\mbl(\Q^n)=\mbl_f(\Q^n)$. 
For any  $M\in\mbl(\Q^n)$ one has $\Q^n=M+M^\perp$. Hence for each $\q\in\Q^n$ there is a unique decomposition $\q=\q_1+\q_2$,
with $\q_1\in M, \q_2\in M^\perp$.  This entails that the map $P_M:\Q^n\to\Q^n$ defined by $P_M\q=\q_1$  is linear, idempotent and Hermitian, that is, $f(\q,P_M\p)=f(P_M\q,\p)$ for all $\q,\p\in\Q^n$. 
For any atom $[\bv]\in\mbl(\Q^n)$, one may thus define the map  $\alpha_{[\bv]}$, with 
\begin{equation}\label{Qprobs}
\alpha_{[\bv]}(M) 
= \frac{f(\bv,P_M\bv)}{f(\bv,\bv)},
\end{equation}
which is  a probability measure on $\mbl(\Q^n)$ and its support is the defining atom, that is, $s(\alpha_{[\bv]}) =[\bv]$.
Clearly, the mapping $\alpha_{[\bv]}\mapsto s(\alpha_{[\bv]})$ gives  a one-to-one correspondence between the set of probability measures on $\mbl(\Q^n)$ of the form $\alpha_{[\bv]}$
and the set of atoms of  $\mbl(\Q^n)$. 

Let $\ps_{\rm at}$ be the set of states defined by the atoms of $\mbl(\Q^n)$, that is, $\alpha\in \ps_{\rm at}$ if $\alpha=\alpha_{[\bv]}$ for some $\bv\in\Q^n,\bv\ne 0$. Any $\sigma$-convex combination of states
$(\alpha_{[\bv_i]})_{i\geq 1}$ with weights $(\lambda_i)_{i\geq 1}$ is again a state (probability measure) on  $\mbl(\Q^n)$.
We let $\sat$ denote the set of all such states. It has
all the regularity properties of Section~\ref{sect:furtherspecifications}, 
including the strong ordering on $\mbl(\Q^n)$.
First of all, each
$\alpha\in\sat$ has a support in $\mbl(\Q^n)$; if $\alpha =\sum_i\lambda_i\alpha_{[\bv_i]}$, then
$s(\alpha)=\lor\{[\bv_i]\,|\, \lambda_i\ne 0\}$. Moreover, if $s(\alpha)=[\bv]$ for some atom $[\bv]$, then $\alpha=\alpha_{[\bv]}$.
Secondly, 
${\rm ex}(\sat)=\ps_{\rm at}$,
which also confirms that the restriction of the support projection to $\ps_{\rm at}$  defines a bijection between the sets ${\rm ex}(\sat)$ and ${\rm At}(\mbl(\Q^n))$.

Let 
${\mathbf S}$ 
denote the set of all probability measures on $\mbl(\Q^n)$. 
We demonstrate next that $\sat$ is a proper subset of ${\mathbf S}$.

To begin with,
we note first that $\mbl(\Q^n)$ can be naturally embedded in $\mbl(\R^n)$. Indeed, for $M\in\mbl(\Q^n)$,
choose an orthogonal basis ${e_1,\ldots,e_k}$, $k\leq n$, with $M={\rm span}_{\Q}\{e_1,\ldots,e_k\}$, and define 
$\widetilde M={\rm span}_{\R}\{e_1,\ldots,e_k\}$. Then $\mbl(\Q^n)\ni M\mapsto \widetilde M\in\mbl(\R^n)$ is an injective mapping.

Fix a nonzero vector $\bv \in\R^n$ such that at least one of its components is irrational. For  $M\in\mbl(\Q^n)$ define
\begin{equation}\label{realextension}
\alpha_{[\bv]}(M)=\frac{\ip{\bv}{P_{\widetilde M}\bv}}{\ip{\bv}{\bv}},
\end{equation}
where $\ip{\cdot}{\cdot}$ denotes the natural inner product in $\R^n$.
 Clearly, $\alpha_{[\bv]}$ is a probability measure on $\mbl(\Q^n)$. However, for any $\q\in\Q^n, \q\ne 0,$
$\alpha_{[\bv]}([\q])\ne 0$, which shows that $\alpha_{[\bv]}$ has not support in  $\mbl(\Q^n)$. Hence $\alpha_{[\bv]}$ is not in $\sat$.

This example shows that the subspace lattice $\mbl(\Q^n)$ of the non-Hilbertian orthomodular space $(\Q^n,\Q,{}^{\rm id},f)$ admits a rich subset of  states ${\sat}$ 
that  has all the listed regularity properties. Therefore, one could  consider $(\sat,\mbl(\Q^n))$ as a logic of a proper quantum system. 
In this case the logic admits also additional probability measures which 
cannot be considered as states of the quantum system since they are not supported in $\mbl(\Q^n)$ .



One may speculate whether $(V,K,{}^*,f)$  can be forced to be a classical
Hilbert space by requiring that the set of all pure probability measures
is defined on $\mbl_f(V)$ so as to be in one-to-one correspondence with the
atoms of $\mbl_f(V)$. 
%
%
Although this seems to be an appealing property, it remains a conjecture   for now, or rather,  a hope for the future.


\section{The role of symmetries in the representation theorem}\label{sec:symmetries}
In his authoritative review \cite{Holland1995} Holland formulated the axiom of {\em ample unitary group} according to which for each pair of mutually orthogonal vectors $u,v\in V$ there is a bijective linear map $U:V\to V$ such that $U(v)=u$ and $f(Ux,Uy)=f(x,y)$ for each $x,y\in V$.
Clearly, this assumption does the job. However,
this is a very strong assumption, and,  in any case, it is not a property given by the pair $(\mbs,\mbl)$. 
Rather than accepting this postulate we follow \cite{CassinelliLahti_2012} to
elucidate the physical content hidden in such an axiom.

\subsection{Implementing symmetries as operators on $V$}

The idea of symmetry receives its natural mathematical representation as a transformation on the set of entities the symmetry refers to. The basic structures are now coded in the sets $\mbl$ and $\mbs$ and in the duality between them. These sets possess various physically relevant structures 
which define the corresponding automorphism groups. From the outset any of them could be used to formulate the notion of symmetry in quantum logic.
In view of the theorem of Sol\'er we shall consider only two of them: symmetries of the set $\atl$ of  atoms of $\mbl$, and the symmetries of the logic $\mbl$.\footnote{Various definitions of the notion of symmetry in quantum mechanics are studied e.g. in
\cite{Gianni_etal2004,Molnar2007}.}

\begin{definition}\label{def:symmetry}
a) A mapping $\ell_o:\atl\to\atl$ is an {\em orthosymmetry}
if it is bijective and  for any $p,q\in \atl$, 
$$
p\perp q\ \Longleftrightarrow\ \ell_o(p)\perp\ell_o(q).
$$
b) A  mapping  $\ell:\mbl\to\mbl$ is a {\em symmetry} if it is bijective and it preserves the order and the orthocomplementation, that is, for any $a,b\in\mbl$,
\begin{eqnarray*}
&&a\leq b \ \Longleftrightarrow\ \ell(a)\leq \ell(b)\\
&&\ell(a^\perp)=\ell(a)^\perp.
\end{eqnarray*}
\end{definition}

Let ${\rm Aut}_o(\atl)$ and  ${\rm Aut}(\mbl)$ denote the sets of orthosymmetries and symmetries  on $\mbl$, respectively. Both of these sets are  groups with respect to the composition of mappings. Moreover, any symmetry $\ell$, when restricted to $\atl$, defines an orthosymmetry.

Assume now that the logic $\mbl$ allows a vector space coordinatization of the form $\mbl\simeq \mbl_f(V)$ for an orthomodular space $(V,K,{}^*,f)$. Any $\ell_o\in{\rm Aut}_o(\atl)$ as well as $\ell\in{\rm Aut}(\mbl)$ defines the corresponding automorphism on the set 
$\ps(V)$ of atoms of $\mbl_f(V)$ and on the whole $\mbl_f(V)$, respectively.
We continue to denote them as $\ell_o$ and $\ell$ and call them orthosymmetry and symmetry, respectively. 

\begin{lemma}\label{FromTSAQM}
Let $\ell_o\in{\rm Aut}_o(\ps(V))$. There is a unique symmetry $\hat\ell_o\in{\rm Aut}(\mbl_f(V))$ such that $\hat\ell_o([v])=\ell_o([v])$ for all $[v]\in\ps(V)$. Moreover, the map ${\rm Aut}_o(\ps(V))\ni\ell_o\mapsto\hat\ell_o\in{\rm Aut}(\mbl_f(V))$ is a group isomorphism.
\end{lemma}

\noindent {\bf Proof.}
Let $\ell_o\in{\rm Aut}_o({\rm At}(\mbl_f(V)))$ and define for any 
(nonempty) subset $M\subseteq V$,
$M\ne\{0\}$,
$$
\hat\ell_o(M)=\{x\in\ell_o([v])\,|\, v\in M, v\ne 0\}
$$
and put $\hat\ell_o(\{0\})=\{0\}$.
Since ${\rm Aut}_o({\rm At}(\mbl_f(V)))$ is a group we also have $\widehat{\ell_o^{-1}}$ defined in the same way. A direct computation shows that
$\hat\ell_o(\widehat{\ell_o^{-1}}(M))= KM$ 
and
$\widehat{\ell_o^{-1}}(\hat\ell_o(M))= KM$. 
Indeed, for any (nonempty) $M$, $M\ne\{0\}$, we have
\begin{eqnarray*}
\widehat{\ell_o^{-1}}(\hat\ell_o(M)) &&= \{w\in \ell^{-1}_0([v])\,|\, v\in \hat{\ell_o}(M)\}\\
&&=\{w\in \ell^{-1}_0([v])\,|\, v\in\ell_o([x]), x\in M, x\ne 0\}\\
&&= \{w\in \ell^{-1}_0(\ell_o([x]))\,|\, x\in M, x\ne 0\}\\
&&=\{w\in[x]\,|\, x\in M, x\ne 0\} = KM,
\end{eqnarray*}
and if $M=\{0\}$, then $\widehat{\ell_o^{-1}}(\hat{\ell_o}(\{0\})=\{0\}$. Similarly, one gets the other set equality.
If $M$ is a subspace, then 
$$
\hat\ell_o(\widehat{\ell_o^{-1}}(M))=M\, ,\quad
\widehat{\ell_o^{-1}}(\hat\ell_o(M))= M.
$$
Using the fact that
for any two (nonzero) vectors $u,v\in V$
 and for any $\ell_o\in {\rm Aut}_o({\rm At}(\mbl_f(V)))$, 
 $f(u,v)=0$  is equivalent to $[u]\perp [v]$ and to $\ell_o([u])\perp \ell_o([v])$ (meaning that
 $f(x,y)=0$ for any $x\in\ell_o([u]), y\in\ell_o([v])$) one easily verifies that
 $$
 \hat{\ell_o}(M)^\perp=\hat{\ell_o}(M^\perp)
 $$ 
 for any (nonempty) set $M\subset V$.
 
 Let now $M\in\mbl_f(V)$. Since $M=(M^\perp)^\perp$, we have
 $\hat{\ell_o}(M)=\hat{\ell_o}(M^\perp)^\perp$ so that by \cite[Lemma 4.35]{Varadarajan1985}   
 $\hat{\ell_o}(M)$ is an $f$-closed subspace of $V$, that is,
 $\hat{\ell_o}(M)\in\mbl_f(V)$. Hence, the map $\mbl_f(V)\ni M\mapsto\hat{\ell_o}(M)\in\mbl_f(V)$ is well-defined. Clearly, it is a bijection, with the inverse  $(\hat\ell_o)^{-1}= \widehat{\ell_o^{-1}}$, it preserves the orthocomplementation and, 
by construction, also the order.
 Therefore, for any $\ell_o\in{\rm Aut}_o({\rm At}(\mbl_f(V)))$, $\hat\ell_o\in{\rm Aut}(\mbl_f(V))$.

For any $v\in V, v\ne 0$, $\hat{\ell_o}([v])=\ell_o([v])$, which shows that $\hat{\ell_o}$ extends the map $\ell_o$. Let $\ell\in{\rm Aut}(\mbl_f(V))$ and assume that 
it is another extension of $\ell_o$.
Since the lattice $\mbl_f(V)$ is atomistic we 
now have 
\begin{eqnarray*}
\ell(M)&=&\ell(\lor\{[v]\,|\, [v]\subseteq M\})=\lor\{\ell([v])\,|\,[v]\subseteq M\}\\
&=& \lor\{\ell_o([v])\,|\,[v]\subseteq M\}
=\lor\{\hat{\ell_o}([v])\,|\,[v]\subseteq M\}\\
&=& \hat{\ell_o}(\lor\{[v]\,|\, [v]\subseteq M\})=\hat{\ell_o}(M)
\end{eqnarray*}
for any $M\in\mbl_f(V)$, showing that 
$\ell=\hat\ell_o$, that is, 
the extension 
is unique.
The map ${\rm Aut}_o({\rm At}(\mbl_f(V)))\ni\ell_o\mapsto\hat\ell_o\in{\rm Aut}(\mbl_f(V))$ is thus  well-defined. Its injectivity and
surjectivity are obvious and it also preserves the group structure:
$\widehat{\ell_o^{-1}}=(\hat{\ell_o})^{-1}$ and 
 $\widehat{\ell_o\circ\ell_o'}=\hat{\ell_o}\circ\hat{\ell_o'}$ for all $\ell_o,\ell_o'\in {\rm Aut}_o({\rm At}(\mbl_f(V)))$.\qed

\ 

Let $\mbl(V)$ be the (complete, irreducible, 
modular, AC) lattice of all subspaces of $V$ and let ${\rm Aut}(\mbl(V))$
be the group of order isomorphisms on $\mbl(V)$. Let $\mbf(V)=\{L\in\mbl(V)\,|\, \dim(L)<\infty\}$ and recall that $\ps(V)\subseteq\mbf(V)\subseteq\mbl_f(V)$. Note also that any $M\in\mbl(V)$ can be expressed as $M=\cup\{L\in\mbf(V)\,|\, L\subseteq M\}=\lor \{L\in\mbf(V)\,|\, L\subseteq M\}$, and $\ell(L)\in\mbf(V)$ for any $L\in\mbf(V), \ell\in{\rm Aut}(\mbl_f(V))$.

\begin{lemma}\label{Fillmore}
For any  $\ell\in{\rm Aut}(\mbl_f(V))$
$$
\Phi_\ell(M)=\cup\{\ell(L)\,|\,  L\subseteq M, L\in\mbf(V)\}
$$
defines an order-preserving bijection $\Phi_\ell:\mbl(V)\to\mbl(V)$ which extends the map $\ell$.
\end{lemma}

\noindent {\bf Proof.} This is an adoptation of the proof of \cite[Lemma 1]{Fillmore1984}.
We show first that $\Phi_\ell(M)\in\mbl(V)$ for any $M\in\mbl(V)$. Indeed, if $x\in\Phi_\ell(M)$, then $x\in\ell(L)$ for some $L\in\mbf(V), L\subseteq M$, and thus
$\lambda x\in\ell(L)\subseteq\Phi_\ell(M)$ for any $\lambda\in K$. Moreover, if $y\in\Phi_\ell(M)$, then $y\in\ell(H)$ for some $H\in\mbf(V), H\subseteq M$, and thus
\begin{eqnarray*}
x+y&\in&\ell(L)+\ell(H)=\ell(L)\lor\ell(H)\\
&=&\ell(L\lor H)
=\ell(L+H)\subseteq\Phi_\ell(M),
\end{eqnarray*}
since the subspaces involved  are all finite dimensional. Hence $\Phi_\ell(M)\in\mbl(V)$ for any $M\in\mbl(V)$.

To prove that the map $\Phi_\ell$ has an inverse, we need the following observation:
$$
\{H=\ell(L)\,|\, L\in\mbf(V), L\subseteq M\} = \{H\in\mbf(V)\,|\, H\subseteq\Phi_\ell(M)\}.
$$
Since $\ell\in{\rm Aut}(\mbl_f(V))$ is a group, 
we also have 
$$
\Phi_{\ell^{-1}}(M)=\cup\{\ell^{-1}(L)\,|\,  L\subseteq M, L\in\mbf(V)\},
$$
 and
using the above observation one quickly confirms that
$$
\Phi_{\ell^{-1}}(\Phi_\ell(M)) = \Phi_{\ell}( \Phi_{\ell^{-1}}(M))
$$
for any $M\in\mbl(V)$. Hence, for any $\ell\in{\rm Aut}(\mbl_f(V))$ the map $\Phi_\ell:\mbl(V)\to\mbl(V)$ is a bijection, with the inverse $\Phi_{\ell^{-1}}$.
By definition  the map $\Phi_\ell$  preserves the order, that is, $\Phi_\ell(M)\subseteq\Phi_\ell(N)$, if and only if $M\subseteq N$ for any $M,N\in\mbl(V)$.
Hence, $\Phi_\ell\in {\rm Aut}(\mbl(V))$ for any $\ell\in  {\rm Aut}(\mbl_f(V))$

We show next that $\Phi_\ell$ extends $\ell$. Let $M\in\mbl_f(V)\subseteq\mbl(V)$. Since
$\Phi_\ell(M)=\cup\{\ell(L)\,|\, L\in\mbf(V), L\subseteq M\}$, $\ell(M)=\cup\{H\,|\, H\in\mbf(V),H\subseteq\ell(M)\}$,
and $H\subseteq\ell(M)$ if and only if $H=\ell(F), F\in\mbf(V),F\subseteq M$, we observe that $\Phi_\ell(M)=\ell(M)$.
Since any $M\in\mbl(V)$ can be expressed as $M=\lor\{L\,|\, L\in\mbf(V), L\subset M\}$ one easily verifies that $\Phi_\ell$ is the only
order isomorphism of $\mbl(V)$ which extends $\ell\in{\rm Aut}(\mbl_f(V))$. \qed

\ 

Let $S:V\to V$ be a bijective map which is $g$-linear, that is, $S$ is bijective and additive on $V$, $g$ an isomorphism of $K$, and $S(\lambda v)= g(\lambda)Sv$ for all $v\in V$ and $\lambda\in K$. Such an $S$ induces an order isomorphism on $\mbl(V)$ by $\Phi_S(M)=\{Sv\,|\, v\in M\}$, and if $S'$ is another bijective $h$-linear map $V\to V$ inducing the same order isomorphism, that is, $\Phi_S=\Phi_{S'}$, then there is a $\lambda\in K$ such that $S\,v=\lambda S'\,v$ for any $v\in V$ \cite[Proposition III.1.2, Corollary III.1.2]{Baer1952}. The first fundamental theorem of projective geometry \cite[p. 44]{Baer1952} gives the converse result: if $\dim(V)\geq 4$, 
then for  any $\Phi\in{\rm Aut}(\mbl(V))$ 
there is an isomorphism $g:K\to K$ and a bijective $g$-linear map $S:V\to V$ inducing $\Phi$.

Let $\ell_o\in{\rm Aut}_o(\ps(V))$ and 
let $S_\ell:V\to V$ thus be a bijective $g_\ell$-linear map which induces the extension 
$\Phi_\ell\in{\rm Aut}(\mbl(V))$ of the extension $\ell\in {\rm Aut}(\mbl_f(V))$ of $\ell_o$. 
Thus, for any $v\in V, v\ne 0$,
$$
S_\ell(Kv)=\Phi_\ell(Kv)=\ell(Kv)=\ell_o(Kv).
$$
Since $\ell_o$ preserves the orthogonality of atoms, the one-dimensional subspaces $S_\ell(Ku)$ and $S_\ell(Kv)$ are orthogonal if and only if $Ku$ and $Kv$ are orthogonal, that is,  $f(Ku,Kv)=0$ if and only if $f(S_\ell(Ku),S_\ell(Kv))=0$.

A direct computation shows that the map 
$$
(u,v)\mapsto g^{-1}_\ell(f(S_\ell u,S_\ell v)) =:\tilde f(u,v)
$$
is a Hermitian form on $(V,K,{}^*)$ such that $\tilde f(u,v)=0$ if and only if $f(u,v)=0$ for all $u,v\in V$.
By virtue of the infinite dimensional version of the Birkhoff - von Neumann theorem \cite{Maeda} there is a nonzero $\rho_\ell\in K$ such that
$\tilde f(u,v)=\rho_\ell f(u,v)$ for all $u,v\in V$.
Moreover, $\rho_\ell$ is a symmetric element of $K$, and since
$\lambda\mapsto\lambda^*$ is an antiautomorphism of $K$ one also has $\lambda\rho_\ell=\rho_\ell\lambda$ for all $\lambda\in K$, that is,
$\rho_\ell\in{\rm Cent}(K)$, the centre of $K$. 


\begin{corollary}\label{prewigner}
For any $\ell_o\in{\rm Aut}_o(\ps(V))$ there is an isomorphism $g_\ell:K\to K$ and a bijective orthogonality preserving $g_\ell$-linear map $S_\ell:V\to V$ such that 
\begin{equation}\label{BvN0}
\ell_o(Kv)=S_\ell(Kv) 
\end{equation}
for any $v\in V$, $v\ne 0$. Moreover, there is a $\rho_\ell\in{\rm Cent}(K) $, $\rho_\ell\ne 0$, $\rho_\ell=\rho^*_\ell$, 
such that
\begin{equation}\label{BvN}
f(S_\ell u,S_\ell v)
= g_\ell(\rho_\ell)g_\ell(f(u,v))
\end{equation}
for all $v,u\in V$.
\end{corollary}

This
corollary is a precursor of the theorem of Wigner according to which the `transition probability preserving bijections on the set of pure states' are implemented by unitary or antiunitary operators acting on the underlying Hilbert space  $\hi$ of the standard logic $\mbl=\mbl(\hi)$ \cite{Varadarajan1985}. 
In that frame, the orthosymmetries are exactly the transition probability
zero  preserving bijections on the pure states. If $\dim(\hi)\geq 3$ then this group coincides with the group of transition probability preserving bijections on the set of pure states \cite[Corollary 4]{Gianni_etal2004}. Now the length of $\mbl$ is at least 4 so that $\dim(V)\geq 3$.

\subsection{Symmetries and the Sol\'er conditions}

We now  study the role of symmetry in providing  a partial justification of the assumptions of Sol\'er's theorem.
 Clearly, the result is obtained if $\mbl\simeq\mbl_f(V)$ has the following property: Given any two mutually orthogonal atoms $[x],[y]\in\mbl_f(V)$, there are  nonzero vectors $x'\in[x]$ and $y'\in[y]$ such that
\begin{equation}\label{ehto}
f(x',x')=f(y',y').
\end{equation}

Before investigating the conditions the theorem of Sol\' er imposes on the set of symmetries, we recall that a proper quantum object is an {\em elementary} quantum object with respect to a group $G$ of (for instance, space-time) transformations if there is a group homomorphism $\sigma:G\to {\rm Aut}_o(\ps(V))$ 
and  if for any pure state (atom) $[v]\in\ps(V)$,
the set $\{\sigma_g([v])\,|\ g\in G\}$  of pure states (atoms) is complete in the sense of superpositions, that is, any other pure state (atom) $[u]\in\ps(V)$ can be expressed as a superposition of some of the pure states (atoms) $\sigma_g([v])$, $g\in G$. Even though this does not solve our problem, it shows that for an elementary quantum object the set of symmetries  ${\rm Aut}_o(\ps(V))$
is rather large and the notion of superposition has a role in it. The next lemma binds the above condition (\ref{ehto}) more tightly to the issue at hand.


\begin{lemma}\label{propositio}
Let $[x],[y]$ be any two mutually orthogonal atoms in $\mbl_f(V)$. If there are nozero vectors $x'\in[x]$ and $y'\in[y]$ such that
$f(x',x')=f(y',y')$ then there is an $\ell_o\in{\rm Aut}_o(\ps(V))$ which swaps $[x]$ and $[y]$, that is, $\ell_o([x])=[y]$ and $\ell_o([y])=[x]$.
Moreover, there is a $[v]\leq [x]\lor [y]$ such that $\ell_o([v])=[v]$.
\end{lemma}
\noindent {\bf Proof.}
Let $M=[x]\lor[y]=[x]\oplus[y]$. Clearly, $[x]=[x'], [y]=[y'].$ Any $u\in M$ can be written uniquely as $u=\alpha x'+\beta y'$, $\alpha,\beta\in K$. Fix $\lambda\in{\rm Cent}(K), \lambda\ne 0$, and define 
$$
U_M(u)= U_M(\alpha x'+\beta y')=\lambda(\alpha y'+\beta x').
$$
The map $U_M$ is a linear bijection on $M$, and for any $u,v\in M$, $\lambda f(u,v)\lambda^* =f(U_Mu,U_Mv)$. Let $\ov=x'+y'$ and observe that $[\ov]$ is fixed by $U_M$. Since $M$ is $f$-closed, $V=M+M^\perp$, so that any  $w\in V$ can uniquely be decomposed as $w=w_1+w_2$, with $w_1\in M, w_2\in M^\perp$. We define a canonical extension of $U_M$ to the whole $V$ by $Uw=U(w_1+w_2)=U_Mw_1+\lambda w_2$. Then $U$ is a bijective linear map on $V$. Moreover, $f(Uw,Uv)=\lambda f(w,v)\lambda^*$ for all $w,v\in V$, and for each $u\in M$, $Uu=U_Mu$. Hence, in particular, 
$$
\Phi_U([x])=[y],\quad \Phi_U([y])=[x],\quad \Phi_U([\ov])=[\ov].
$$
\qed
\ 

This lemma shows that condition (\ref{ehto}) implies the existence of a special symmetry of $\mbl_f(V)$ that interchanges the two orthogonal atoms $[x]$ and $[y]$ and has
a superposition of them as a fixed point.

To get the opposite implication, and thus come to the final conclusion, 
we add the following two assumptions, the first concerning the group ${\rm Aut}_o(\ps(V))$, the second
the form $f$: 

\

\begin{itemize}
\item[] ({\bf A})\ \ 
The symmetry group is {\em abundant} in the following sense:
for any pair of mutually orthogonal atoms $[x],[y]\in\ps(V)$ there is a symmetry $\ell_o\in{\rm Aut}_o(\ps(V))$ that swaps $[x]$ and $[y]$, that is,
$\ell_o([x])=[y]$ and $\ell_o([y])=[x]$, 
and has some of their
superpositions as a fixed point, that is, $\ell_o([v])=[v]$ for some $[v]\leq[x]\lor[y]$;

\

\item[] ({\bf R})\ \ 
The form $f$ is {\em regular} in the following sense:
for each $v\in V$,
$f(v,v)\in {\rm Cent}(K)$, and 
$g(f(v,v))=f(v,v)$ for any automorphism  $g$ of  $K$.
\end{itemize}

\begin{lemma}\label{toinen}
Let $[x],[y]$ be any two mutually orthogonal atoms in $\mbl_f(V)$. If
the group ${\rm Aut}_o(\ps(V))$ is abundant
and
the form $f$ is regular then there are nonzero vectors $x'\in[x]$ and $y'\in[y]$ such that $f(x',x')=f(y',y')$. 
\end{lemma}

\noindent {\bf Proof.} 
Let $\ell_o\in{\rm Aut}_o(\ps(V))$ be an orthosymmetry swapping the  atoms $[x]$ and $[y]$ and having a $[v]\leq [x]\lor[y]$ as a fixed point.
Let $S_\ell,g_\ell,\rho_\ell$ constitute a realization of $\ell_o$ as given in Corollary~\ref{prewigner}. Applying Eq. (\ref{BvN}) first to the vector $v$ and its transform $S_\ell v=\lambda v$, $\lambda\in K$,
one gets $g_\ell(\rho_\ell)=\lambda\lambda^*$. Applying then the same equation to $x$ and $S_\ell x=\alpha y, \alpha\in K$, one gets
$$
f(\alpha y,\alpha y)
=
g_\ell(\rho_\ell) g_\ell(f(x,x)) =\lambda f(x,x)\lambda^*
=f(\lambda x,\lambda x)
$$
which completes the proof. \qed

\ 

We summarize the results of this section in the form of a theorem.

\begin{theorem}\label{thm:Soler}
Assume that the logic $(\mbs,\mbl)$ of the statistical duality $(\mbs,\mbo,p)$  has the structure of Corollary~\ref{AClattice1}. 
Assume that the  system 
has an abundant set of orthosymmetries. 
If there is an infinite sequence of mutually orthogonal atoms in $\mbl$, 
and if the form $f$ of the coordinatization $(V,K,{}^*,f)$ of the logic is regular,
then $V$ is a Hilbert space over $\R,\C$, or $\quat$, and $\mbl$ is (ortho-order) isomorphic with the lattice of closed subspaces of the Hilbert space $V$.
\end{theorem}

With this theorem 
the statistical duality $(\mbs,\mbo,p)$ of a proper quantum system is completely resolved:
the states $\alpha\in\mbs$ of the system are identified with 
positive trace one operators $\rho$ of an infinite dimensional classical Hilbert space $\hi$, 
the observables $(\sfe,\Omega,\hA)\in\mbo$ are expressed as semispectral measures, also called normalized positive operator measures, taking values in the set of bounded operators on $\hi$,
and the numbers $p(\alpha,\sfe,X)$ are determined to be given by the `Born rule' $p(\alpha,\sfe,X)=\tr{\rho\sfe(X)}$. The pure states are  the one-dimensional projections and the $\mbl$-valued observables are the spectral measures.
 
We are left with the question whether
the regularity  of the form $f$, the requirement ({\bf R}), can be stated as a property of the logic $(\mbs,\mbl)$ of the duality $(\mbs,\mbo,p)$.
Another open question is the choice of 
the number field left open by Theorem~\ref{Solertheorem}. We close our paper with a short comment on this.

It is  well known that the complex Hilbert space $\hi$ is in many respects simpler than the real or quaternionic Hilbert spaces. 
We recall only the powerfull polarization identity (valid in the complex case) and the formulation of the Stone theorem which is of fundamental importance. But is the choice $\C$ only a mathematical convenience? Some of the differences between  the three cases have been discussed already in \cite[Chapter 22]{EnricoGianni}. 
In addition to that we mention here  the work of Pulmannov\' a \cite{Pulmannova1996} where a symmetry argument is given to rule out the real and quaternionic choices for $K$.
Finally we note that the cases of Hilbert spaces over $\C$ and
$\R$ can be distinguished in terms of the different lower bounds obtained
in the respective derivations of Heisenberg-Kennard-Robertson -type preparation uncertainty relations 
\cite{LahtiMaczynski1987}.

\subsection*{Acknowledgement} We are grateful to Drs Paul Busch and  Maciej M\c aczynski for  their valuable comments in earlier versions of this manuscript.


\begin{thebibliography}{99}
\bibitem{Alfsen1971} E.M. Alfsen, {\em Compact Convex Sets and Boundary Integrals}, Springer, Berlin, 1971.
\bibitem{Baer1952} R. Baer, {\em Linear Algebra and Projective Geometry}, Academic Press, New York, 1952.
\bibitem{EnricoGianni} E. Beltrametti, G. Cassinelli, {\em The Logic of Quantum Mechanics}, Addison-Wesley, Reading, Massachusetts, 1981.
Cambridge University Press, 1985, 2010.
\bibitem{BvN1936} G. Birkhoff, J. von Neumann,
The logic of quantum mechanics,
{\em Annals of mathematics} {\bf 37} (1936) 823-843. 
\bibitem{Bohr1935} N. Bohr,  Can quantum-mechanical description of physical reality be considered complete? {\em Physical Review} {\bf 48} (1935) 696-702.
\bibitem{BB1973} K.Bugajska, S. Bugajski, The projection postulate in quantum logic, {\em Bull. Ac. Pol. Sci. Series des sciences math., astr. et phys.} {21} (1973) 873-877.
\bibitem{BugLah1980} S. Bugajski, P. Lahti, Fundamental principles of quantum theory, {\em Int. J. Theor. Phys.} {\bf 19} (1980) 499-514.
\bibitem{CassinelliBeltrametti1975} G. Cassinelli, E. Beltrametti, Ideal, first-kind measurements in a proposition-state structure, {\em Commun. math. Phys.} {\bf 40} (1975) 7-13.
\bibitem{Gianni_etal2004} G. Cassinelli, E. De Vito, P.J. Lahti, A. Levrero, {\em The Theory of Symmetry Actions in Quantum Mechanics},
Springer, {\bf LNP 654}, 2004.
\bibitem{CassinelliLahti_2012} G. Cassinelli, P. Lahti, A theorem of Sol\'er, the theory of symmetry, and quantum mechanics,
{\em Int. J. Geometric Methods in Modern Physics}, {\bf 9} (2012) 1260005(7).
\bibitem{Chiribellaetal2010} G. Chiribella, G.M. D'Ariano, P. Perinotti, Probabilistic theories with purification, {\em Physical Review A} {\bf 81} (2010) 062348(40).
\bibitem{Chiribellaetal2011} G. Chiribella, G.M. D'Ariano, P. Perinotti, Informational derivation of quantum theory, {\em Physical Review A} {\bf 84} (2011) 012311(39).
\bibitem{Clifton_etal2003} R. Clifton, J. Bub, H. Halvorson, Characterizing quantum theory in terms of information-theoretic constrains, arXiv:quant-ph/0211089v2 19 Feb 2003.
\bibitem{CornetteGudder1974} W.M. Cornette, S.P. Gudder, The mixture of quantum states, {\em J. Math. Phys.} {\bf 15} (1974) 842-850.
\bibitem{DaviesLewis1970} E.B. Davies, J.L. Lewis, An operational approach to quantum probability, {\em Commun. math. phys.} {\bf 17} (1970) 239-260.
\bibitem{Davies1976} E.B. Davies, {\em Quantum Theory of Open Systems}, Academic Press, London, 1976.
\bibitem{Dirac1958} P.A.M. Dirac, {\em The Principles of Quantum Mechanics}, Oxford University Press, London 1958. First edition 1930.
\bibitem{Edwards1970} C.M. Edwards, The operational approach to algebraic quantum theory I, {\em Commun. math. Phys.} {\bf 16} (1970) 207-230.
\bibitem{Edwards1971} C.M. Edwards, Classes of operations in quantum theory, {\em Commun. math. Phys.} {\bf 20} (1971) 26-36.
\bibitem{Edwards1972} C.M. Edwards, The theory of pure operations, {\em Commun. math. Phys.} {\bf 24} (1972) 260-288.
\bibitem{EdwardsGerzon1970} C.M. Edwards, M.A. Gerzon, Monotone convergence in partially ordered vector spaces, {\em Ann. Inst. Henri Poincer\'e} {\bf 12} (1970) 323-328..
\bibitem{Ellis1964} A.J. Ellis, The duality of partially ordered normed linear spaces, {\em Journal London Math. Soc.} {\bf 39} (1964) 730-744.
\bibitem{Ellis1966} A.J. Ellis, Linear operators in partially ordered normed vector spaces, {\em J. London Math. Soc.} {\bf 41} (1966) 323-332.
\bibitem{Fillmore1984} F.A. Fillmore, W.E. Longstaff, On isomorphisms of lattices of closed subspaces, {\em Can. J. Math.} {\bf XXXVI} (1984) 820-829.
\bibitem{FoulisRandall1973} D.J. Foulis, C.H. Randall, The empirical logic approach to the physical sciences, pp 230-249 in \cite{HartkamperNeumann1974}.
\bibitem{FoulisRandall1974} D.J. Foulis, C.H. Randall, Empirical logic and quantum mechanics, {\em Synthese} {\bf 29} (1974) 81-111.
\bibitem{Gleason1957} A.M. Gleason, Measures on the closed subspaces of a {H}ilbert space, {\em J. Math. Mech.} {\bf 6} (1957)
885--893.
\bibitem{GrossKunzi} H. Gross, U.-M. K\"unzi, On a class of orthomodular quadratic spaces, {\em L'Enseignement Math\'ematique} {\bf 31} (1985) 187-212.
\bibitem{Gudder1973} S.P. Gudder, Convex structures and operational quantum mechanics, {\em Commun. Math. Phys.} {\bf 29} (1973) 249-264.
\bibitem{Gudder1979} S.P. Gudder, {\em Stochastic Methods in Quantum Mechanics}, Elsevier North Holland, 1979.
\bibitem{HartkamperNeumann1974} A. Hartk\"amper, H. Neumann, eds., {\em Foundations of Quantum Mechanics and Ordered Linear Spaces}, Springer, Berlin, 1974.
\bibitem{Holland1995} S.S. Holland, Orthomodularity in infinite dimensions; a theorem of M. Sol\' er, {\em Bull. Am. Math. Soc.} {\bf 32} (1995) 205-234.
\bibitem{Jauch1968} J.M. Jauch, {\em Foundations of Quantum Mechanics}, Addison-Wesley, 1968.
\bibitem{JauchPiron1963} J.M. Jauch, C. Piron, Can hidden variables be excluded in quantum
mechanics? {\em Helvetica Physica Acta} {\bf 36} (1963) 827-837.
\bibitem{Keller} H. Keller, Ein nicht-klassischer Hilbertscher Raum, {\em Math. Z.} {\bf 172} (1980) 41-49.
\bibitem{Keller1984} H. Keller, Measures on non-classical Hilbertian spaces, {\em Notas mathematicas} {\bf No 16}, Universidad Catoliga Santiago, Chile 1984, 49-71.
\bibitem{BugLah1985} P. Lahti, S. Bugajski,  Fundamental principles of quantum theory. II. From a convexity scheme to the DHB theory. {\em Int. J. Theor. Phys.} {\bf 24} (1985) 1051-1080.
\bibitem{LahtiMaczynski1987} P.J. Lahti, M. J. M\c aczynski, Heisenberg inequality and the complex field in quantum mechanics, {\em J. Math. Phys.} {\bf 28} (1987) 1764-1769.
\bibitem{LahtiMaczynski1992}  P.J. Lahti, M. J. M\c aczynski, Orthomodularity and quadratic transformations in probabilistic theories of physics,
{\em J. Math. Phys.} {\bf 33} (1992) 4133-4138.
\bibitem{Loomis} L.H. Loomis, On the representation of $\sigma$-complete Boolean algebras, {\em Bulletin of the American Mathematical Society} {\bf 35} (1947) 757-760.
\bibitem{Ludwig1967} G. Ludwig, Attempt of an axiomatic foundation of quantum mechanics and more general theories (II), {\em Commun. math. phys} {\bf 4} (1967) 331-348. 
\bibitem{Ludwig1983} G. Ludwig, {\em Foundations of Quantum Mechanics I}, Springer, 1983.
\bibitem{Mackey1963} G. Mackey, {\em Mathematical Foundations of Quantum Mecianics}, W.A. Benjamin Inc., New York, 1963. 
\bibitem{Maczynski1973} M. J. M\c aczynski, The orthogonality postulate in axiomatic quantum mechanics, {\em Int. J. Theor. Phys.} {\bf 8} (1973) 353-360.
\bibitem{Maeda} F. Maeda, S. Maeda, {\em Theory of Symmetric Lattices}, Springer, Berlin, 1970.
\bibitem{Mielnik1969} B. Mielnik, Theory of filters, {\em Commun. math. phys.} {\bf 15} (1969) 1-46.
\bibitem{Mielnik1974} B. Mielnik, Generalized quantum mechanics, {\em Commun. math. phys.} {\bf 37} (1974) 115-152.
\bibitem{Mittelstaedt1978} P. Mittelstaedt, {\em Quantum Logic}, D. Reidel Publ. Co., Dordrecht, 1978.
\bibitem{Molnar2007} L. Moln\' ar, {\em Selected Preserver Problems in Algebraic Structures of Linear Operators and on Function Spaces}, Springer, {\bf LNM 1895}, 2007.
\bibitem{Morash} R.P. Morash, Angle bisection and orthoautomorphisms in Hilbert lattices, {\em Can. J. Math.} {\bf 25} (1973) 261-272.
\bibitem{Piron1976} C. Piron, {\em Foundations of Quantum Physics}, Benjamin, Reading, Massachusetts, 1976.
\bibitem{Piziak} R. Piziak, Orthomodular lattices and quandratic spaces: a survey, {\em Rocky Mountain Journal of Mathematics} {\bf 21} (1991) 951-992.
\bibitem{Pulmannova1996} S. Pulmannov\'a, Axiomatization of quantum logics, {\em Int. J. Theor. Phys.} {\bf 35} (1996) 2309-2319.
\bibitem{Pool1968} J.C.T. Pool, Baer${}^*$-semigroups and the logic of quantum mechanics, {\em Commun. math. Phys.} {\bf 9} (1968) 118-141.
\bibitem{Pool1968b} J.C.T. Pool, Semimodularity and the logic of quantum mechanics, {\em Commun. math. Phys.} {\bf 9} (1968) 212-228.
\bibitem{PavelSylvia} P. Pt\'ak, S. Pulmannov\'a, {\em Orthomodular Structures as Quantum logics}, Kluwer, Dordrecht, 1991.
\bibitem{Redei1998} M. R\' edei, 
{\em  Quantum Logic in Algebraic Approach},  
{\em Fundamental Theories of Physics}  {\bf 91},
Kluwer Academic Publishers, Dordrecht, Boston  and London, 1998. 
\bibitem{Schaefer1971} H.H. Schaefer, {\em Topological Vector Spaces}, Springer. Berlin, 1971.
\bibitem{Schaefer1974} H.H. Schaefer,  Orderings of vector spaces, pp 4-10 in \cite{HartkamperNeumann1974}. 
\bibitem{Soler1995} M.P. Sol\'er, Characterization of Hilbert spaces by orthomodular spaces, {\em Communications in Algebra} {\bf 23} (1995) 219-243.
\bibitem{Sikorski} R. Sikorski, {\em Boolean Algebras}, Springer, Berlin, 1964.
\bibitem{Stone} M.H. Stone, The theory of representations of Boolean algebras, {\em Transactions of the American Mathematical Society} {\bf 40} (1936) 37-111.
\bibitem{Stone1949} M.H. Stone, Postulates for the barycentric calculus, {\em  Ann. Mat. Pura Appl.} {\bf 29} (1949), 25–30. 
\bibitem{Varadarajan1985} V.S. Varadarajan, {\em Geometry of Quantum Theory}, 2nd edition, Springer, 1985.
\bibitem{Wilbur} W.J. Wilbur, On characterizing the standard quantum logic, {\em Trans. Am. Math. Soc.} {\bf 233} (1977) 265-281.
\bibitem{Zierler1961} N. Zierler, Axioms for non-relativistic quantum mechanics, {\em Pacific J. Math.} {\bf 11} (19961) 1151-1169.

\end{thebibliography}
\end{document}